\documentclass[12pt]{article}
\usepackage{bm}
\usepackage{graphicx}
\usepackage{verbatim, color, array}
\usepackage{hyperref}
\usepackage{amsthm, amssymb}
\usepackage[affil-it]{authblk} % Package for affiliation
\usepackage{float}
\usepackage{hhline}
\usepackage{booktabs}
\usepackage{soul}
\usepackage[table]{xcolor}
\usepackage{multirow}
\usepackage{nicematrix}
\usepackage{threeparttable}
\usepackage{enumitem}
\usepackage{tikz}
\usepackage[super]{natbib}
\usepackage{setspace}
\usetikzlibrary{shapes, arrows, positioning, calc, shadows, fit}
\usepackage{longtable}
\usepackage{fancyhdr}
\usepackage[margin=1in,textwidth=6.5in]{geometry}
\title{Statistical Operating Characteristics of Current Early Phase Dose Finding Designs with Toxicity and Efficacy in Oncology}
\date{}
\author[1]{Hao Sun}
\author[1]{Hsin-Yu Lin}
\author[2]{Jieqi Tu}
\author[1]{Revathi Ananthakrishnan}
\author[1]{Eunhee Kim \thanks{Corresponding author: *Eunhee Kim, Global Biometrics \& Data Sciences, Bristol Myers Squibb, New Jersey, USA. Email: \href{mailto:Eunhee.Kim@bms.com}{Eunhee.Kim@bms.com}}}
\affil[1]{Global Biometrics \& Data Sciences, Bristol Myers Squibb, New Jersey, USA}
\affil[2]{Division of Epidemiology and Biostatistics, School of Public Health, University of Illinois at Chicago, Illinois, USA}

\begin{document}
\doublespacing
\maketitle

\vspace{-0.5in}
\textbf{Abstract:} Traditional phase I dose finding cancer clinical trial designs aim to determine the maximum tolerated dose (MTD) of the investigational cytotoxic agent based on a single toxicity outcome, assuming a monotone dose-response relationship. However, this assumption might not always hold for newly emerging therapies such as immuno-oncology therapies and molecularly targeted therapies, making conventional dose finding trial designs based on toxicity no longer appropriate. To tackle this issue, numerous early phase dose finding clinical trial designs have been developed to identify the optimal biological dose (OBD), which takes both toxicity and efficacy outcomes into account. In this article, we review the current model-assisted dose finding designs, BOIN-ET, BOIN12, UBI, TEPI-2, PRINTE, STEIN, and uTPI to identify the OBD and compare their operating characteristics. Extensive simulation studies and a case study using a CAR T-cell therapy phase I trial have been conducted to compare the performance of the aforementioned designs under different possible dose-response relationship scenarios. The simulation results demonstrate that the performance of different designs varies depending on the particular dose-response relationship and the specific metric considered. Based on our simulation results and practical considerations, STEIN, PRINTE, and BOIN12 outperform the other designs from different perspectives.


\newpage

\section{Introduction}
Finding the right dose is crucial in early phase cancer clinical trials due to the safety concerns of adverse events induced by any oncology drug under investigation. There are numerous real world examples where the dose on the label of an approved oncology drug is not the dose administered in practice \cite{markman2010serious, minasian2014optimizing}. This implies that early phase dose finding oncology trials need to be carefully designed to determine the optimal dose of the investigational agent that can be used in late phase trials and the target population. In this regard, many early phase oncology dose finding methods have been proposed in the past few years.  

The current existing dose finding methods can be categorized into algorithmic or rule-based, model-assisted, and model-based designs \cite{yuan2019model}. The 3+3 design is an example of a rule-based design, which is commonly used because of its simplicity and ease of implementation. However, due to the limitations of the 3+3 design \cite{araujo2021contemporary}, many model-assisted and model-based designs were proposed to address them. For example, some popular model-assisted dose finding designs include the modified toxicity probability interval (mTPI) design \cite{ji2010modified}, the mTPI-2 design \cite{li2017toxicity, yan2017keyboard, pan2020keyboard}, and the Bayesian optimal interval (BOIN) design \cite{liu2015bayesian, yuan2016bayesian}. Commonly used model-based dose finding designs include the continual reassessment method (CRM) design \cite{iasonos2011continual}, the escalation with overdose control (EWOC) design \cite{babb1998cancer}, and the Bayesian logistic regression model (BLRM) design \cite{neuenschwander2008critical}. All these designs only consider the toxicity of the drug to estimate the maximum tolerated dose (MTD). This works well for chemotherapies where both the efficacy and toxicity of these agents increase with dose.  

However, traditional dose finding designs based on a single toxicity outcome may not be appropriate for immuno-oncology drugs and molecularly targeted therapies. For these novel anticancer agents, although the toxicity of the drug increases with dose, the efficacy of the drug may not always increase with dose. Additionally, in many immunotherapies, toxicities are usually low or moderate grade, preventing the observation of a dose-limiting toxicity (DLT). Also, severe toxicities are rare and often delayed in subsequent treatment cycles, which may prevent the MTD from being reached \cite{postel2011phase, weber2015toxicities}. Hence, it is important to identify the dose of such an oncology drug that is efficacious as well as tolerable, which is referred to as the optimal biological dose (OBD). Recognizing the importance of selecting the right dose, the FDA has recently launched the initiative Project Optimus \cite{us2022project} to reform the dose optimization and dose selection paradigm in oncology drug development. There has been an ongoing effort to develop early phase clinical trial designs incorporating both toxicity and efficacy outcomes to determine the OBD. Model-based dose finding designs aimed at OBD detection include the Eff-Tox design \cite{thall2004dose}, the local logistic model (L-logistic) design \cite{zang2014adaptive}, the change-point logistic (CP-logistic) model design \cite{sato2016adaptive}, and the Bayesian dynamic model (B-dynamic) design \cite{liu2016robust}. Eff-Tox \cite{thall2004dose} is an adaptive model-based design based on trade-offs between the treatment efficacy and toxicity probabilities. It is particularly sensitive to the efficacy-toxicity trade-off contour. The L-logistic design \cite{zang2014adaptive} utilizes Bayesian logistic regression to model the dose-response curve. The CP-logistic method\cite{sato2016adaptive} introduces a change‐point logistic model to account for the correlation between efficacy and toxicity outcomes.  The B-dynamic method \cite{liu2016robust} leverages a flexible Bayesian dynamic model that considers both toxicity and efficacy and borrows information across different dose levels. However, a limitation of the B-dynamic method is that it considers only monotone dose-response relationships.

Furthermore, an increasing number of model-assisted designs extend BOIN and mTPI by accounting for efficacy outcomes in the dose finding procedures. Some popular extensions of the BOIN design are BOIN-ET \cite{takeda2018boin}, BOIN12 \cite{lin2020boin12}, UBI \cite{li2020tepi}, and STEIN \cite{lin2017stein}, while extensions of the mTPI design include TEPI \cite{li2017toxicity}, TEPI-2 \cite{li2020tepi}, PRINTE \cite{lin2021probability}, and uTPI \cite{shi2021utpi}. BOIN-ET derives optimal probability intervals for efficacy and toxicity by minimizing the joint probabilities of incorrect dosing decisions to determine the OBD. UBI, BOIN12, and uTPI use utility functions to assess the toxicity-efficacy trade-off and select the dose with the highest utility as the OBD. Unlike UBI, which bases dosing decisions on data from the current dose, BOIN12 and uTPI adaptively compare multiple doses to choose the optimal dose for the next group of patients. TEPI, TEPI-2, and UBI designs rely on pre-specified decision tables that map two-dimensional toxicity and efficacy probability intervals to a set of dosing decisions. On the other hand, uTPI proposes a chessboard design that utilizes a two-dimensional square grid of toxicity and desirability to guide dosing decisions.

Multiple model-assisted designs mentioned above, such as BOIN-ET, BOIN12, and PRINTE, demonstrated that they are more robust than the Eff-Tox design as they do not make any model assumptions regarding the dose toxicity and efficacy curves. BOIN-ET was shown a higher OBD selection probability than CP-logistic \cite{takeda2018boin}. Lin \& Yin \cite{lin2017stein} observed that STEIN outperforms both L-logistic and B-dynamic in OBD selection percentage. TEPI-2, UBI, PRINTE, BOIN-ET, and STEIN demonstrated superior performance compared to TEPI in terms of OBD selection probability and patient allocation at the optimal doses \cite{li2020tepi, lin2021probability, yamaguchi2023optimal}. Yamaguchi et al.\cite{yamaguchi2023optimal} presented evidence that BOIN-ET and STEIN have similar OBD selection probabilities. Li et al. \cite{li2023comparison} found that BOIN-ET performs better than BOIN12 in more than half of the scenarios. Shi et al. \cite{shi2024comparative} compared BOIN-ET, BOIN12, TEPI, PRINTE, Joint3+3, STEIN, uTPI, and Eff-Tox by applying the same utility score approach for OBD selection across all designs. They revealed that BOIN12 and uTPI outperformed other designs in OBD selection accuracy and minimizing poor dose allocation. Additionally, STEIN demonstrated the most effective overdose control. However, as discussed by Shi et al., the designs, except for BOIN12 and uTPI, can underperform when the OBD is defined based on the utility score method \cite{shi2024comparative}. Furthermore, they did not consider scenarios where the true OBD does not exist. 

In this article, we compare seven model-assisted early phase dose finding oncology designs: BOIN-ET, BOIN12, TEPI-2, UBI, PRINTE, STEIN, and uTPI. Several model-based designs, including Eff-Tox, L-logistic, CP-logistic, and B-dynamic, as well as the model-assisted design TEPI, are not discussed in this article because their performance has been revealed to be less competitive compared to one or more of the aforementioned model-assisted designs. To the best of our knowledge, the performance of the selected seven designs, using their original dose exploration algorithms and OBD estimation approaches, has not been fully examined in the literature. The goal of our article is to provide comprehensive guidance on selecting an appropriate early phase design under different scenarios in practice and to offer user-friendly, interactive software that can be used to implement these designs.

This article is organized as follows. Section~\ref{sec2_notation} introduces some basic design parameters, OBD definitions, and safety and futility rules. Key concepts of different early phase dose finding designs are described in Section~\ref{sec3_methods}. Extensive simulation studies and sensitivity analyses are presented in Section~\ref{simu_sec}, and a case study is featured in Section~\ref{case_sec}. Section~\ref{sec_soft} provides the design guidance and software information. Further discussion is provided in Section~\ref{discussion}.

\vspace{-0.2in}
\section{Notations}\label{sec2_notation}
\vspace{-0.15in}
Consider an early phase trial with binary toxicity and efficacy outcomes and a total of $D$ dose levels. The true toxicity probability and efficacy probability for the $d$-th dose level are defined as $p_d$ and $q_d$, respectively, $d = 1, \ldots, D$, where $p_d$ and $q_d$ are assumed to be independent. The toxicity probabilities are assumed to be strictly increasing, i.e. $p_1 < \ldots < p_D$. However, the efficacy probabilities are assumed to have an unknown dose-response relationship. From Lin et al.\cite{lin2020boin12}, the combination of a binary toxicity outcome and a binary efficacy outcome consists of four different outcomes: 1 = (no toxicity, efficacy); 2 = (no toxicity, no efficacy); 3 = (toxicity, efficacy); 4 = (toxicity, no efficacy). The number of patients of the $i$-th outcome under dose level $d$ is denoted by $y_{d,i}$, $i = 1, 2, 3, 4$. Let $y_{d,T} = y_{d,3} + y_{d,4}$ and $y_{d, E} = y_{d, 1} + y_{d, 3}$ be the number of patients experiencing a DLT and efficacy response under dose level $d$, respectively, and $n_d = \sum_i^4 y_{d, i}$ be the total number of patients assigned to dose level $d$. We further define $\pi_{d,i}$ and $\hat{\pi}_{d, i} = y_{d,i}/n_d$ as the true probability and the observed probability of the $i$-th outcome, $\hat{p}_d$ as the observed toxicity rate, and $\hat{q}_d$ as the observed efficacy rate at dose level $d$. Then, $p_d = \pi_{d, 3} + \pi_{d, 4}$, $q_d = \pi_{d, 1} + \pi_{d, 3}$, $\hat{p}_d= y_{d,T}/n_d$, and $\hat{q}_d = y_{d,E}/n_d$. Assume that the maximum number of cohorts in the trial is $C_M$, where each cohort represents a group of patients joining the trial over a certain period. Let the sample size of each cohort be $n_c$, so that the maximum total sample size of the trial can be expressed as $N = C_Mn_c$.

\vspace{-0.2in}
\subsection{Optimal Biological Dose}\label{obd}

The maximum acceptable toxicity probability and the minimum acceptable efficacy probability are defined as $p_T$ and $q_E$, respectively. The MTD is the dose level $d_{MTD}$ which has the largest toxicity probability not greater than $p_T$, i.e. $d_{MTD} = \arg\max_{d} p_d I(p_d \leq p_T)$, which can be any dose level among the $D$ available dose levels. The MTD may not exist if all dose levels have a toxicity rate greater than $p_T$. Compared to the MTD, the OBD can be defined using one of three different methods:  

(1) \textbf{Utility Score}: The utility score approach, employed by BOIN12 and uTPI, defines the OBD by assigning different utility values, $\{u_i\}_{i=1}^4$, to the combinations of toxicity and efficacy outcomes, as shown in Table~\ref{utility2d}. This method offers flexibility and can be extended to ordinal toxicity and efficacy outcomes. The OBD is defined as the dose yielding the highest expected utility calculated as $EU_d = \sum_{i = 1}^4 \pi_{d,i}u_i$, with acceptable toxicity and efficacy probabilities,  i.e. 
\vspace{-0.1in}
$$
d^{US}_{OBD} = \arg\max_d EU_dI(p_d\leq p_T, q_d\geq q_E).
\vspace{-0.1in}
$$

(2) \textbf{Utility Function}: The utility function approach involves a predefined utility function $U(p_d, q_d)$ based on toxicity and efficacy probabilities. This method is widely adopted in designs such as STEIN, TEPI-2, UBI, and PRINTE, although the specific utility functions may vary. The OBD is the dose level that maximizes this function with acceptable toxicity and efficacy probabilities, i.e. 
\vspace{-0.2in}
$$
d^{UF}_{OBD} = \arg\max_d U(p_d, q_d)I(p_d\leq p_T, q_d\geq q_E).
\vspace{-0.2in}
$$

(3) \textbf{Maximum Efficacy}: The maximum efficacy approach, adopted in BOIN-ET, selects the dose with the highest  efficacy probability and acceptable toxicity probability given by
\vspace{-0.2in}
$$
d^{ME}_{OBD} = \arg\max_d q_dI(p_d\leq p_T, q_d\geq q_E).
$$
\vspace{-0.05in}
When multiple doses share the highest utility value or efficacy probability, the OBD is the one with the lowest toxicity rate. If no dose level satisfies the criterion of having both acceptable toxicity and efficacy probabilities, specifically $p_d \leq p_T$ and $q_d \geq q_E$, then the true OBD does not exist. The maximum efficacy approach represents a specific case of both the utility score and utility function approaches. In the utility score approach, by setting $u_2 = u_4 = 0$ and $u_1 = u_3 > 0$, the expected utility simplifies to $EU_d = u_1q_d$, thereby designating the OBD as the dose level with the highest efficacy probability. Besides, in the utility function method, if the utility function, $U(q_d)$, depends only on the efficacy probability, then the OBD is the dose with the highest efficacy probability. 

% a special case of the utility functions, $U(q_d)$, depends only on the efficacy probability and defines the OBD as the dose with the highest efficacy probability. 

With the true toxicity and efficacy probabilities, each design has its own suggested OBD, which is the true OBD defined by each specific design, given its pre-specified design parameters. We adopted the same OBD definition approach as applied in the original paper for each design to ensure that the suggested OBD in our paper was consistent with the one defined in its original paper. In our simulation, we also ensured that the randomly chosen true OBD aligned with the suggested OBD of each design at the start of each replication. If a design had a different suggested OBD, we repeated the generation of true toxicity and efficacy probabilities until the alignment was achieved across all designs. A detailed discussion is provided in the supplementary document.

% (revision) ensuring that the randomly chosen OBD aligned with the preferred OBD as determined by all dose finding designs given the true toxicity and efficacy probabilities. In this paper, we utilized the same utility score values and utility functions as suggested in the original papers of these dose finding designs. In our simulation studies, the dose level as the true OBD is randomly selected in each replication. However, we ensure that the randomly selected dose level is consistently as the true nominated OBD for all designs. } Because BOIN-ET employs the maximum efficacy approach which is the special case of the other two approaches, we } Considering that BOIN-ET employs the maximum efficacy approach, we repeated the generation procedure using the maximum efficacy method until the OBD prioritized by all designs consistently matched the same chosen dose level.

% A good design is defined as a design with a high probability of selecting the OBD when the OBD exists and terminating the trial without selecting any dose if the OBD does not exist. 

% the set of possible OBD is defined as $\{0, 1, \ldots, D\}$ where 0 means that no dose level is considered as the OBD. 

\vspace{-0.2in}
\subsection{Safety and Futility Rules}
A dose level $d$ is considered admissible if the observed data indicate that dose level $d$ is reasonably safe and efficacious, i.e. $p_d \leq p_T$ and $q_d \geq q_E$. After assigning each cohort of patients to a specific dose level, we check the following safety and futility rules: 

\newlength{\mylength}
\settowidth{\mylength}{\textbf{(Futility)}}
\begin{itemize}[left=0em]
    \item[] \makebox[\mylength][l]{\textbf{(Safety)}} \parbox[t]{0.9\linewidth}{if $\Pr(p_d > \phi_T\mid y_{d, T}, n_d) > \eta$, eliminate the current and all above doses from the dose list;}
    \item[] \makebox[\mylength][l]{\textbf{(Futility)}} \parbox[t]{0.9\linewidth}{if $\Pr(q_d < \phi_E\mid y_{d,E}, n_d) > \xi$, eliminate the current dose from the dose list.}
\end{itemize}
Only admissible doses can be used to treat patients. If there are no admissible doses to assign the next cohort of patients to, then the trial will be terminated. Different dose finding designs use different values for the parameters $\{\phi_T, \phi_E, \eta, \xi\}$. Specifically, $\{\phi_T, \phi_E, \eta, \xi\}$ equals to $\{0.35, 0.25, 0.95, 0.9\}$ for BOIN12, $\{0.4, 0.2, 0.9, 0.9\}$ for BOIN-ET, $\{0.33, 0.3, 0.95, 0.98\}$ for STEIN, $\{0.3, 0.25, 0.95, 0.9 \}$ for uTPI, and $\{0.4, 0.2, 0.9, 0.7\}$ for TEPI-2, UBI, and PRINTE. In this paper, we set $\{\phi_T, \phi_E, \eta, \xi\} = \{0.35, 0.25, 0.95, 0.8\}$ across all the designs to ensure a fair comparison among all designs. 

% The hyperparameters  $\{\phi_T, \phi_E, \eta, \xi\}$ are specified differently in each design in Section~\ref{sec3_methods}. For example, $\{\phi_T, \phi_E, \eta, \xi\}$ equals to $\{0.35, 0.25, 0.95, 0.9\}$ for BOIN12, $\{0.4, 0.2, 0.9, 0.9\}$ for BOIN-ET, and $\{0.4, 0.2, 0.9, 0.7\}$ for TEPI-2, UBI, and PRINTE.

% {For example, BOIN12 sets $\{\phi_T, \phi_E, \eta, \xi\}$ equal to $ \{0.35, 0.25, 0.95, 0.9\}$, BOIN-ET sets $\{\phi_T, \phi_E, \eta, \xi\} = \{0.4, 0.2, 0.9, 0.9\}$, TEPI-2, UBI, and PRINTE set $\{\phi_T, \phi_E, \eta, \xi\} = \{0.4, 0.2, 0.9, 0.7\}$}. 

\section{Methods}\label{sec3_methods}
\vspace{-0.1in}
\subsection{BOIN-ET}
BOIN-ET directly extends the idea of BOIN by utilizing both binary toxicity and efficacy outcomes \cite{takeda2018boin}. In BOIN-ET, there are two cut points on the toxicity interval, $\lambda_1$ and $\lambda_2$, and one cut point, $\eta_1$, on the efficacy interval, which slices the combination of toxicity and efficacy intervals into 6 regions as illustrated in Table~\ref{BOIN-ET_dec}. BOIN-ET determines the optimal values for $(\lambda_1, \lambda_2, \eta_1)$ under the restriction $\phi_1 < \lambda_1 < \phi_p < \lambda_2 < \phi_2$ and $\delta_1 < \eta_1 < \delta_E$, where $\phi_p$ and $\delta_E$ denote the target toxicity probability and efficacy probability, respectively, where $\phi_1 = 0.1 p_T$, $\phi_2 = 1.4 p_T$, and $\delta_1 = 0.6\delta_E$. Takeda et al. \cite{takeda2018boin} developed 6 composite hypotheses and the following decision table as shown in Table~\ref{BOIN-ET_dec}. The values of these cut points are selected through a grid search to minimize the joint probability of incorrect dosing decisions. When $0 \leq \hat{p}_d \leq \lambda_1$ and $\eta_1 < \hat{q}_d \leq 1$ are observed, the next cohort of the patients will be dosed at the current dose level using BOIN-ET, while BOIN chooses a higher dose level to dose the next cohort. In BOIN-ET, if $\lambda_1 < \hat{p}_d < \lambda_2$ and $0\leq \hat{q}_d \leq \eta_1$, escalation, staying at the same dose, and de-escalation are all possible choices due to the unknown dose-response relationship. On the contrary, BOIN only considers staying at the current dose if $\lambda_1 < \hat{p}_d < \lambda_2$. Let $c = 1$ denote the initial cohort and $d = 1$ the pre-specified initial dose level. The dose finding algorithm of BOIN-ET follows: 

\vspace{-0.1in}
\begin{itemize}
    \item[1. ] Treat the cohort $c$ at the dose level $d$. 
    \vspace{-0.1in}
    \item[2. ] Calculate $\hat{p}_d = y_{d,T}/n_d$ and $\hat{q}_d = y_{d, E}/n_d$ at the current dose level $d$ and follow the decision table in Table~\ref{BOIN-ET_dec} to make the decision of escalation/stay/de-escalation. Specifically, if $\lambda_1 < \hat{p}_d < \lambda_2$ and $0\leq \hat{q}_d \leq \eta_1$, define the admissible dose set $A_d = \{d - 1, d, d + 1\}$ and consider the following cases: 
    \vspace{-0.1in}
    \begin{itemize}
            \item[(a) ] if dose level $d + 1$ has never been used before, escalate to $d + 1$;
            \vspace{-0.1in}
            \item[(b) ] if (a) is not applicable, choose the admissible dose that has the maximum probability of efficacy according to $\hat{q}_{d-1}$, $\hat{q}_{d}$, and $\hat{q}_{d+1}$;   
            \vspace{-0.1in}
            \item[(c) ] if (a) and (b) are not applicable because at least 2 doses have the same maximum probability of efficacy, randomly choose 1 dose among the doses that exhibit the maximum probability of efficacy. 
            \vspace{-0.1in}
    \end{itemize}
    \vspace{-0.1in}
    \item[3. ] Set $c = c+1$ and update $d$ based on the decision in Step 2. 
    \vspace{-0.1in}
    \item[4. ] Repeat Steps 1 - 3 until the maximum total sample size is reached. 
\end{itemize}

\vspace{-0.1in}
BOIN-ET first applies isotonic regression on $\{\hat{p}_d\}_{d=1}^D$ to determine the MTD. For efficacy, BOIN-ET applies fractional polynomial regression with 2 degrees of freedom to estimate the efficacy probabilities which allows a non-monotonic dose-response relationship. The model deviance is used to select the best-fitting model containing 2 powers $(k_1, k_2)$ from the set $\{-2, -1, -0.5, 0, 0.5, 1, 2, 3\}$. Among the tolerable doses, the OBD is determined by the best-fitting polynomial model.

\vspace{-0.1in}
\subsection{BOIN12}
BOIN12 is a flexible model-assisted design to find the OBD that optimizes the risk-benefit trade-off \cite{lin2020boin12}. BOIN12 defines a utility $u_i$ for each outcome $i$ as shown in Table~\ref{utility2d}. The best outcome (no toxicity, efficacy) and the worst outcome (toxicity, no efficacy) have the utility values $u_1 = 100$ and $u_4 = 0$, respectively. The other two outcomes have utility values $u_2$ and $u_3$ within the range $[0, 100]$. As specified in Section~\ref{sec2_notation}, BOIN12 utilizes the expected utility to select the OBD, which is the dose level $d$ with the largest mean utility $u_d = \sum_{i=1}^4 u_i\pi_{d,i}$. When $u_2 + u_3 = 100$, $u_d = u_2(1-p_d) + u_3q_d$. Moreover, if $u_2 = 0$ and $u_3 = 100$, then $u_d = 100q_d$. In this special case, the OBD is the most efficacious dose level. We assume $u_2 = 40$ and $u_3 = 60$ as recommended by the authors to balance the risk-benefit trade-off. 

Lin et al.\cite{lin2020boin12} introduced a rank-based desirability score (RDS) for the decisions during the trial. The standardized desirability of $d$ is defined as $u^*_d = 100^{-1}\sum_{i=1}^4 u_i\pi_{d,i} \in [0, 1]$, which can be considered as a weighted average of $\{\pi_{d,i}\}_{i=1}^4$. Then suppose $x_d = 100^{-1}\sum_{i=1}^4 u_iy_{d,i}$ which can be interpreted as the number of events observed from $n_d$ patients with event probability $u^*_d$. Under the Bayesian framework, assigning $u^*_d$ a $\text{Beta}(\alpha, \beta)$ prior leads to the posterior distribution of $u^*_d$ given the data to be $u^*_d \mid n_d, x_d \sim \text{Beta}(\alpha + x_d, \beta + n_d - x_d)$. The RDS is the rank of the posterior probabilities of $\Pr(u_d^* > u_b\mid n_d, x_d)$ for all possible situations, where the benchmark $u_b$ for comparison is a constant within the range $[0, 1]$. In other words, the next dose level is selected based on the posterior probabilities of the dose levels in the admissible set. Table 3 of Lin et al.\cite{lin2020boin12} is an example of an RDS Table. Let $c = 1$ denote the initial cohort and $d = 1$ the pre-specified initial dose level. The dose finding algorithm of BOIN12 follows from Figure 1 of Lin et al.\cite{lin2020boin12}:

\vspace{-0.1in}
\begin{itemize}
    \item[1. ] Treat the cohort $c$ at the dose level $d$. 
    \vspace{-0.1in}
    \item [2. ] Calculate the observed DLT rate $\hat{p}_d = y_{d,T}/n_d$ at the current dose level $d$ and compare it with two constants $0 < \lambda_e < \lambda_d <1$, where $\lambda_e$ and $\lambda_d$ are from BOIN \cite{liu2015bayesian}. 
        \begin{itemize}
        \vspace{-0.1in}
            \item[(a) ] if $\hat{p}_{d} \geq \lambda_d$, de-escalate to the next lower dose level $d - 1$; 
            \vspace{-0.1in}
            \item[(b) ] if $\lambda_e < \hat{p}_d < \lambda_d$: (i) when $n_d \geq 6$, select the dose from $\{d-1, d\}$ with larger RDS; (ii) when $n_d < 6$, select the dose from $\{d-1, d, d+1\}$ with the largest RDS; 
            \vspace{-0.1in}
            \item[(c) ] if $\hat{p}_d \leq \lambda_e$, select the dose from $\{d-1, d, d+1\}$ with the largest RDS.
        \end{itemize}
    \vspace{-0.1in}
    \item[3. ] Set $c = c+1$ and update $d$ based on the decision in Step 2. 
    \vspace{-0.1in}
    \item[4. ] Repeat Steps 1 - 3 until the maximum total sample size is reached. 
\end{itemize}

\vspace{-0.1in}
To prevent the dose finding process from getting stuck at a locally optimal dose, BOIN12 has an additional dose exploration rule: if $\hat{p}_d < \lambda_d$ and $n_d \geq 9$ for the current dose, and the next higher dose level has not been used, escalate to the next higher dose level $d+1$. To select the OBD, there are two following steps: (1) determine the MTD by applying isotonic regression to the observed toxicity rates $\{\hat{p}_d\}_{d=1}^D$ and choosing the dose level with the closest isotonically estimated toxicity rate to $p_T$; (2) select the dose level with the highest estimated utility as the OBD among the doses with acceptable estimated toxicity probabilities. 

\vspace{-0.2in}
\subsection{UBI}
Li et al.\cite{li2020tepi} developed UBI by combining BOIN with a utility function $U(\hat{p}_d, \hat{q}_d) = f_E(\hat{q}_d) - \theta f_T(\hat{p}_d)$ to incorporate efficacy with toxicity, where 
$$
f_E\left(\hat{q}_d\right)=\left\{\begin{array}{c}
0, \quad \hat{q}_d> \theta_{Eff} \\
\hat{q}_d, \quad \hat{q}_d \leq \theta_{Eff}
\end{array}\right., 
\qquad
f_T\left(\hat{p}_d\right)=\left\{\begin{aligned}
& 0, \quad \hat{p}_d \leq \theta_{Tox} \\
& 1, \quad \hat{p}_d \geq \lambda_d \\
& 0, \quad \hat{p}_d \in (\theta_{Tox}, \lambda_e], \text{ and }\hat{q}_d \leq \theta_{Eff}\\
& \hat{p}_d, \quad  \hat{p}_d \in (\lambda_e, \lambda_d) \text{ and } \hat{q}_d \leq \theta_{Eff}\\
& \hat{p}_d/3, \quad  \hat{p}_d \in (\theta_{Tox}, \lambda_d) \text{ and } \hat{q}_d > \theta_{Eff}
\end{aligned}\right.,
$$
$\lambda_e$ and $\lambda_d$ are the same parameters defined in BOIN and $\{\theta, \theta_{Eff}, \theta_{Tox}\} = \{2, 0.66, 0.15\}$ are the three design parameters defined in UBI. Specifically, when $\theta = 2$, the toxicity utility is considered to be twice as important as the efficacy utility.  Let $c = 1$ denote the initial cohort and $d = 1$ be the pre-specified initial dose level. The dose finding algorithm of UBI is as follows: 

\vspace{-0.1in}
\begin{itemize}
    \item[1. ] Treat the cohort $c$ at the dose level $d$. 
    \vspace{-0.1in}
    \item[2. ] Based on the observed data, calculate $\hat{p}_d = y_{d,T}/n_d$ and $\hat{q}_d = y_{d, E}/n_d$, and the utility function $U(\hat{p}_d, \hat{q}_d)$ at the current dose level $d$. 
    \begin{itemize}
    \vspace{-0.1in}
        \item[(a) ] if $U \geq 0$, escalate to the next higher dose; 
        \vspace{-0.1in}
        \item[(b) ] if $U < -1/3$, de-escalate to the next lower dose; 
        \vspace{-0.1in}
        \item[(c) ] otherwise, if $ -1/3 \leq U < 0$, stay at the current dose.
    \end{itemize}
    \vspace{-0.1in}
    \item[3. ] Set $c = c+1$ and update $d$ based on the decision in Step 2. 
    \vspace{-0.1in}
    \item[4. ] Repeat Steps 1 - 3 until the maximum total sample size is reached. 
\end{itemize}
\vspace{-0.1in}
At the end of the trial, the OBD is selected as the dose with the highest utility score with a utility function defined as $U(\tilde{p}_d, \hat{q}_d) = g_E(\hat{q}_d) - \theta g_T(\tilde{p}_d)$, where $\tilde{p}_d$ is the isotonically transformed estimate of the toxicity probability from the observed toxicity probability $\hat{p}_d$, $g_E(\hat{q}_d)$ and $g_T(\tilde{p}_d)$ are truncated functions given by 

$$
g_T\left(\hat{p}_i\right)=\left\{\begin{array}{c}
p_{g1}^*, \hat{p}_i<p_{g1}^* \\
p_{g2}^*, \hat{p}_i>p_{g2}^*, \\
\hat{p}_i, \text { else }
\end{array}\right. \quad
\quad g_E\left(\hat{q}_i\right)=\left\{\begin{array}{c}
q_{g1}^*, \hat{q}_i<q_{g1}^* \\
q_{g2}^*, \hat{q}_i>q_{g2}^*\\
\hat{q}_i, \text { else }
\end{array} \right.
,
$$
and $\{p_{g1}^*$, $p_{g2}^*$, $q_{g1}^*$, $q_{g2}^*\} = \{0.15, 0.4, 0.2, 0.6\}$ are pre-specified cutoff values. 

\vspace{-0.1in}
\subsection{TEPI-2}\label{method_TEPI-2}
Guo et al.\cite{guo2017bayesian} proposed an extension of the mTPI design \cite{ji2010modified}, called the mTPI-2 design, by dividing the toxicity interval into subintervals of equal length. This modification aims to mitigate undesirable decisions that may arise from the mTPI design. Li et al.\cite{li2017toxicity} developed TEPI that extends the idea of the mTPI design by incorporating efficacy into the dose finding model. Similar to mTPI-2, TEPI-2 improves TEPI by dividing both the toxicity and efficacy intervals into subintervals with equal lengths, respectively \cite{li2020tepi}. TEPI-2 partitions the toxicity unit into 4 subintervals, \{Low, Moderate, High, Unacceptable\}, and the efficacy unit into 4 subintervals,\{Low, Moderate, High, Superb\}. The TEPI-2 decision table needs to be pre-determined in consultation with physicians. In the example TEPI-2 decision table shown in Table~\ref{TEPI-2_dec}, the lengths of the toxicity subintervals and efficacy subintervals are 0.08 and 0.2, respectively. Each interval rectangle constructed by a pair of toxicity and efficacy subintervals is assigned a decision. Unlike BOIN-ET, TEPI-2 escalates to a higher dose level with low toxicity but superb efficacy. With the observed data, TEPI-2 computes the joint utility probability mass (JUPM) for each interval rectangle, $(a_1, b_1) \times (a_2, b_2)$ as 
$$
JUPM^{(a_2, b_2)}_{(a_1, b_1)} = \frac{\Pr(p_d \in (a_1, b_1), q_d \in (a_2, b_2) \mid n_d, y_{d, T}, y_{d, E})}{(b_1 - a_1)\times (b_2 - a_2)}, \quad 0 < a_1 < b_1 < 1, 0 < a_2 < b_2 < 1. 
$$
The dosing decision for the next cohort is based on which interval rectangle has the largest JUPM. Let $c = 1$ denote the initial cohort and $d = 1$ the pre-specified initial dose level. The dose finding algorithm of TEPI-2 is as follows: 

\vspace{-0.1in}
\begin{itemize}
    \item[1. ] Treat the cohort $c$ at the dose level $d$. 
    \vspace{-0.1in}
    \item[2. ] Based on the observed data, calculate the JUPMs for all interval rectangles and select the rectangle $(a_1, b_1) \times (a_2, b_2)$ with the largest JUPM. The dosing decision of the next cohort is based on what is given in the PRINTE decision table corresponding to the rectangle with the largest JUMP.
    \vspace{-0.1in}
    \item[3. ] Set $c = c+1$ and update $d$ based on the decision in Step 2. 
    \vspace{-0.1in}
    \item[4. ] Repeat Steps 1 - 3 until the maximum total sample size is reached. 
\end{itemize}
\vspace{-0.1in}
At the end of the trial, the OBD is selected as the dose with the highest estimated posterior expected utility. The utility score function is defined as $U(p, q) = f_1(p)f_2(q)$, where both $f_1(p)$ and $f_2(q)$ are truncated functions, given by
$$
f_1(p) = \left\{\begin{aligned} 
1, \quad & p\in (0, p_1^*] \\ 1 - \frac{p - p_1^*}{p_2^* - p_1^*}, \quad & p \in (p_1^*, p_2^*) \\ 0, \quad & p \in [p_2^*, 1) \end{aligned}\right., \qquad 
f_2(q) = \left\{\begin{aligned} 
0, \quad & q\in (0, q_1^*] \\ \frac{q - q_1^*}{q_2^* - q_1^*}, \quad & q \in (q_1^*, q_2^*) \\ 1, \quad & q \in [q_2^*, 1) \end{aligned}.\right.
$$
The estimated posterior expected utility is given by $\hat{E}[U(p_d, q_d)\mid n_d, y_{d, T}, y_{d, E}] = T^{-1}\sum_{t=1}^T $ $U^t(\hat{p}_d^t, q_d^t)$, where $T$ is the Monte Carlo size, $\{p_d^t, q_d^t\}_{d, t}$ are generated from the posterior distributions of toxicity and efficacy rates, and $\hat{p}_d^t$ is derived after isotonic transformation. 

\vspace{-0.1in}
\subsection{PRINTE}\label{method_PRINTE}

Similar to TEPI-2, PRINTE is an extension of TEPI and divides the efficacy and toxicity intervals into subintervals with equal lengths, respectively. Unlike TEPI-2, the decision table of PRINTE is pre-specified based on the values of the target toxicity rate $p_T$, target efficacy rate $p_E$, and a small fraction value $\epsilon$. Per the PRINTE decision table, the dose needs to be de-escalated if $p_d$ is greater than $p_T + \epsilon$. If $p_d$ is not greater than $p_T + \epsilon$, the next cohort of patients is dosed at the current dose $d$ if $q_d \geq p_d$ or at the next higher dose if $q_d < p_d$. An example PRINTE decision table is presented in Table~\ref{PRINTE_dec}. Let $c = 1$ denote the initial cohort and $d = 1$ the pre-specified initial dose level. The dose finding algorithm of PRINTE is as follows: 
\vspace{-0.1in}
\begin{itemize}
    \item[1. ] Treat the cohort $c$ at the dose level $d$. 
    \vspace{-0.1in}
    \item[2. ] Based on the observed data, calculate the JUPMs for all interval rectangles and select the rectangle $(a_1, b_1) \times (a_2, b_2)$ with the largest JUPM. The decision for the next cohort is the decision of the rectangle with the largest JUPM from the PRINTE decision table. 
    \vspace{-0.1in}
    \item[3. ] Set $c = c+1$ and update $d$ based on the decision in Step 2. 
    \vspace{-0.1in}
    \item[4. ] Repeat Steps 1 - 3 until the maximum total sample size is reached. 
\end{itemize}
\vspace{-0.1in}
Once the maximum total sample size is attained, PRINTE determines the optimal dose by selecting the one with the highest estimated posterior expected utility. Let $A(p, q) = \{(p, q)\mid p\in (0, p_T], q\in [q_E + \delta, 1)\}$ be a graduate region and $p_{in} = \Pr(U(\hat{p}_d^t, \hat{q}_d^t) \in B)$, where $B$ is the corresponding graduate utility region for $(p, q) \in A(p, q)$. The optimal dose will be selected as the OBD if $p_{in} \geq p_{graduate}$ where $p_{graduate}$ denotes a threshold value. Otherwise, the optimal dose will be rejected and the OBD will be considered as non-existent. 

\vspace{-0.1in}
\subsection{STEIN} 
STEIN \cite{lin2017stein} is another extension of BOIN, which utilizes a frequentist model averaging approach to estimate the efficacy probabilities. Similar to BOIN-ET, STEIN also has two cut points, $\psi_L$ and $\psi_U$, on the toxicity intervals, and one cut point, $\phi$, on the efficacy interval. Unlike BOIN-ET, which utilizes a grid search to estimate these cut points, STEIN can directly calculate these cut points using the same approach as BOIN \cite{yuan2016bayesian}. When $\hat{p}_d < \psi_U$ and $\hat{q}_d < \psi$, STEIN also constructs an admissible set for the dosing decision of the next cohort. Let $c = 1$ denote the initial cohort and $d = 1$ the pre-specified initial dose level. The dose finding algorithm of STEIN is as follows: 
\vspace{-0.1in}
\begin{itemize}
    \item[1. ] Treat the cohort $c$ at the dose level $d$. 
    \vspace{-0.1in}
    \item[2. ] Calculate $\hat{p}_d = y_{d,T}/n_d$ and $\hat{q}_d = y_{d, E}/n_d$ at the current dose level $d$. 
    \vspace{-0.1in}
    \begin{itemize}
        \item[(a)] if $\hat{p}_d \geq \psi_U$, escalate to $d + 1$; 
        \vspace{-0.1in}
        \item[(b)] if $\hat{p}_d < \psi_L$ and $\hat{q}_d \geq \phi$, stay at the current dose $d$; 
        \vspace{-0.1in}
        \item[(c)] if $\hat{p}_d \leq \psi_L$ and $\hat{q}_d < \phi$, define the admissible set $A_d = \{d-1, d, d+1\}$ and select the dose $d^{Next}$ for the next cohort with $d^{Next} = \arg\max_{d' \in A_d} \Pr(q_{d'} > \psi\mid n_{d'}, y_{d', E})$; 
        \vspace{-0.1in}
        \item[(d)] if $\psi_L < \hat{p}_d < \psi_U$ and $\hat{q}_d < \phi$, define the admissible set $A_d = \{d-1, d\}$ and select the dose $d^{Next}$ for the next cohort with $d^{Next} = \arg\max_{d' \in A_d} \Pr(q_{d'} > \psi\mid n_{d'}, y_{d', E})$. 
    \end{itemize}
    \vspace{-0.1in}
    \item[3. ] Set $c = c+1$ and update $d$ based on the decision in Step 2. 
    \vspace{-0.1in}
    \item[4. ] Repeat Steps 1 - 3 until the maximum total sample size is reached. 
\end{itemize}
\vspace{-0.1in}
At the end of the trial, STEIN uses isotonic regression on $\{\hat{p}_d\}_{d=1}^D$ to obtain the isotonically transformed values $\{\tilde{p}_d\}_{d=1}^D$. For the efficacy outcomes, STEIN performs $D$ unimodal isotonic regressions on $\{\hat{q}_d\}_{d=1}^D$ by enumerating all possible models in the dose-efficacy curve. In the $d'$th model, $d' = 1, \ldots, D$, the dose level $d'$ attains the highest efficacy probability. With unimodal isotonic transformations, the transformed efficacy probabilities, $\{\tilde{q}_{d'd}\}_{d=1}^D$, have $\tilde{q}_{d'1} \leq \cdots \leq \tilde{q}_{d'd'} \geq \cdots \geq \tilde{q}_{d'D}$. The pseudo-likelihood based on the $d'$th unimodal isotonic regression is given by 
$$
\tilde{L}_{d'} = \prod_{d = 1}^D \binom{n_d}{y_{d, E}}\tilde{q}_{d'd}^{y_{d, E}}(1-\tilde{q}_{d'd}^{y_{d, E}})^{n_j - y_{d,E}}.
$$
The final model averaging estimate of $q_d$ is given by $\tilde{q}_d = \sum_{d' = 1}^D \pi_{d'}\tilde{q}_{d'd}$, where $\pi_{d'} = \tilde{L}_{d'}/\sum_{d=1}^D \tilde{L}_d$. The OBD is the dose level with the highest utility, as defined by the utility function $U(\tilde{p}_d, \tilde{q}_d) = \tilde{q}_d - w_1\tilde{p}_d - w_2\tilde{p}_dI(\tilde{p}_d > p_T)$ \cite{liu2016robust}, where $w_1 = 0.33$ and $w_2 = 1.09$, as recommended in the original paper \cite{lin2017stein}.

\subsection{uTPI}
uTPI combines the ideas of modeling dose desirability and the chessboard design method for dose finding \cite{shi2021utpi}. Similar to BOIN12, uTPI defines the expected utility or desirability as $EU_{d} = \sum_{i = 1}^4 u_i\pi_{d, i}/100$ and the observed utility value as $OU_d = \sum_{i=1}^4 u_iy_{d,i}/100$, where $u_i$ is the utility score of outcome $i$ in Table~\ref{utility2d}. uTPI assumes that the numerical utility outcome has a pseudo binomial distribution so that the quasi-likelihood for $EU_d$ is $L(OU_d \mid EU_d) = EU_d^{OU_d}(1-EU_d)^{n_d - OU_d}$. By assigning a noninformative beta prior $EU_{d} \sim \text{Beta}(1, 1)$, the posterior distribution of $EU_{d}$ is $EU_{d}\mid \text{Data} \sim \text{Beta}(1 + OU_d, 1 + n_d - OU_d)$. Unlike the previous designs, uTPI constructs a two-dimensional chessboard by dividing the joint square of toxicity probability and desirability into equally sized squares. The toxicity interval and desirability interval are divided into subinverals, $\{\mathcal{I}_{k, T}\}_{k = 1}^{10}$ and $\{\mathcal{I}_{k,U}\}_{k=1}^{10}$, where $\mathcal{I}_{k,T} = \mathcal{I}_{k,U} = [0.1(k-1), 0.1k)$ for $k = 1, \ldots, 9$ and $\mathcal{I}_{10,T} = \mathcal{I}_{10,U} = [0.9, 1]$. The strongest toxicity interval and the strongest desirability interval are defined as $k_d^T = \arg\max_k \Pr(p_{d} \in \mathcal{I}_{k,T}\mid \text{Data})$ and $k_d^U = \arg\max_k \Pr(EU_{d} \in \mathcal{I}_{k,U}\mid \text{Data})$, respectively. Let $k^*$ denote the index of the toxicity subinterval such that $p_T \in \mathcal{I}_{k^*, T}$. Let $c = 1$ denote the initial cohort and $d = 1$ the pre-specified initial dose level. The dose finding algorithm of uTPI is as follows: 
\vspace{-0.1in}
\begin{itemize}
    \item[1. ] Treat the cohort $c$ at the dose level $d$. 
    \vspace{-0.1in}
    \item[2. ] Identify the strongest toxicity interval index $k_{d}^T$ and  strongest desirability interval index $k_{d}^U$ for all the dose levels.
    \vspace{-0.1in}
    \begin{itemize}
        \item[(a)] if $k_d^T > k^*$, de-escalate to dose level $d-1$;
        \vspace{-0.1in}
        \item[(b)] if $k_d^T < k^*$, choose dose level $d'$ from the admissible set $\{d - 1, d, d + 1\}$ that has the largest $k_d^U$;
        \vspace{-0.1in}
        \item[(c)] if $k_d^T = k^*$, choose dose level $d'$ from the admissible set $\{d - 1, d, d + 1\}$ if $n_d < N*$ or from the admissible set $\{d - 1, d\}$ if $n_d \geq N*$ that has the largest $k_d^U$.
    \end{itemize}
    \vspace{-0.1in}
    \item[3. ] Set $c = c+1$ and update $d$ based on the decision in Step 2. 
    \vspace{-0.1in}
    \item[4. ] Repeat Steps 1 - 3 until the maximum total sample size is reached. 
\end{itemize}
\vspace{-0.1in}

In Step 2, if there is a tie, uTPI selects the dose level $d'$ with the maximum $\Pr(OU_{d'} > \bar{\mathcal{I}}_{k_{d'}^U, U} \mid \text{Data})$, where $\bar{\mathcal{I}}_{k_{d'}^U, U}$ denotes the upper boundary of the $k_{d'}^U$th desirability subinterval. At the end of the trial, uTPI also applies an isotonic regression on $\{\hat{p}_d\}_{d=1}^D$ to identify $d_{MTD}$, which is the dose level with the estimated toxicity rate $\tilde{p}_{d}$ closest to $p_T$. The OBD is selected as the dose level that has the maximum posterior mean of $EU_d$, i.e. $d_{OBD} = \arg\max_{d \leq d_{MTD}} \widehat{EU}_d$.

\vspace{-0.2in}
\section{Simulation Studies}\label{simu_sec}
\vspace{-0.1in}
\subsection{Simulation Settings} \label{sec:simu_settings}
We conducted simulation studies to compare the operating characteristics of the seven designs introduced in Section~\ref{sec3_methods}. The toxicity probabilities are assumed to increase monotonically with dose levels.  For efficacy probabilities, we introduce four different dose-response relationships as follows:
\newlength{\lengthtwo}
\newlength{\lengththree}
\settowidth{\lengthtwo}{\textbf{Increasing (I): }}
\settowidth{\lengththree}{\textbf{Increasing-Plateau (IP): }}
\vspace{-0.1in}
\begin{itemize}[left=-2em]
\item[] \makebox[\lengthtwo][l]{\textbf{Increasing (I): }} \parbox[t]{0.9\linewidth}{efficacy probabilities increase monotonically with dose levels, i.e. $q_1 < \cdots < q_D$;}
\item[] \makebox[\lengthtwo][l]{\textbf{Constant (C): }} \parbox[t]{0.9\linewidth}{efficacy probabilities are the same across all dose levels, i.e. $q_1 = \cdots = q_D$;}
\item[] \makebox[\lengthtwo][l]{\textbf{Unimodal (U): }} \parbox[t]{0.9\linewidth}{efficacy probabilities increase monotonically until a certain dose level $d^U < D$, \\ after which they decrease monotonically, i.e. $q_1 < \cdots < q_{d^U}$ and $q_{d^U} > \cdots > q_D$;}
\item[] \hspace{-0.8in}\makebox[\lengththree][l]{\textbf{Increasing-Plateau (IP): }} \parbox[t]{1.9\linewidth}{efficacy probabilities increase monotonically until a certain dose level $1 < d^{IP} < D$, \\ after which they remain constant, i.e. $q_1 < \cdots < q_{d^{IP}} = \cdots = q_D$. }
\end{itemize}

\begin{figure} 
    \centering
    \includegraphics[width = 7.5in, height = 5in]{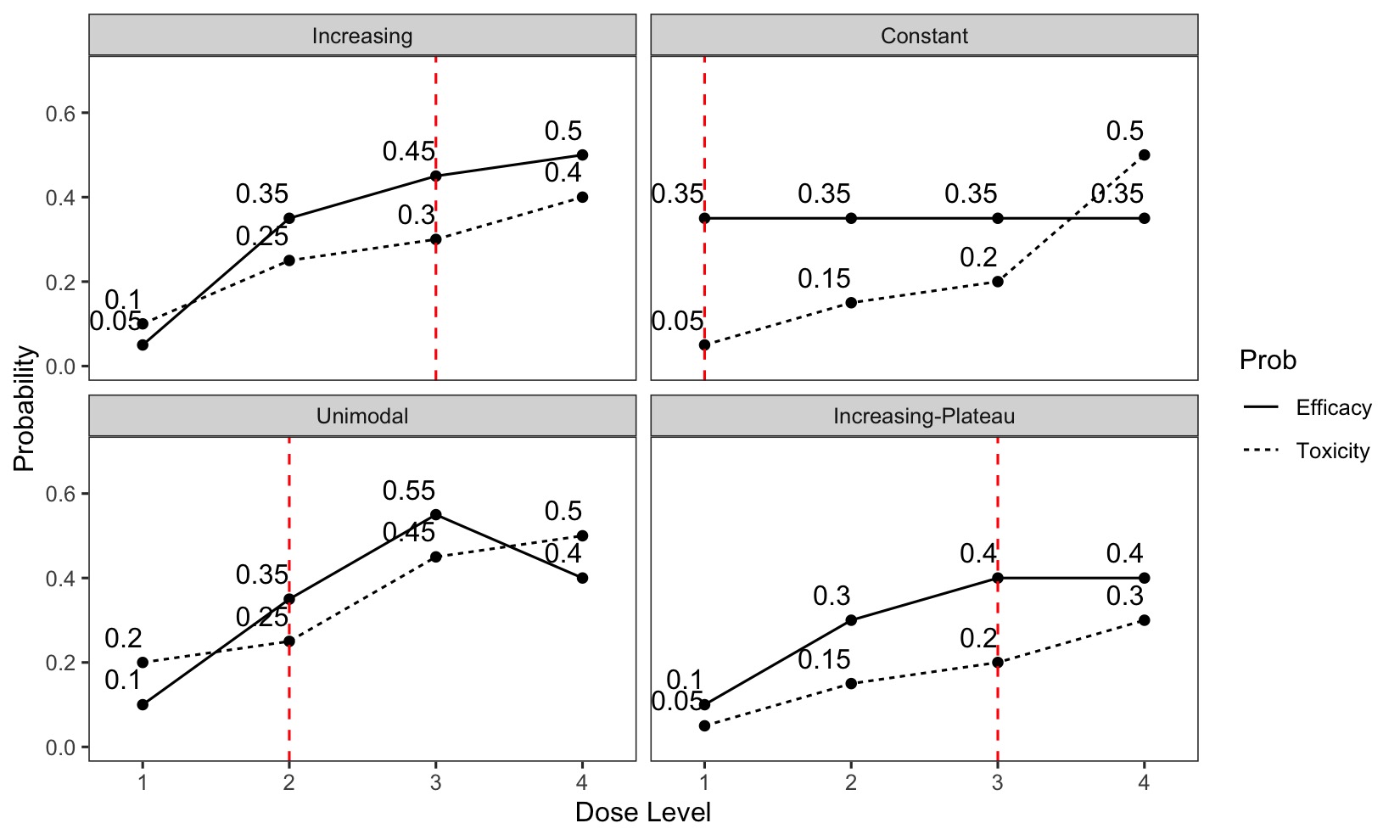}
    \caption{Example toxicity (dashed lines) and efficacy (solid lines) curves with true OBD (red vertical dashed lines) under four different dose-response relationships}
    \label{fig_effcurve}
    \begin{minipage}{\textwidth} 
        \footnotesize
        \vspace{0.2in}
        (1) $d^U = 3$ for Scenario U; (2) $d^{IP} = 3$ for Scenario IP; (3) $(d^*, d^H) = \{(3, 4), (1, 1), (2, 3), (3,3)\}$ for Scenarios I, C, U, and IP, respectively.  
    \end{minipage}
\end{figure}

In the unimodal case, we assumed that $d^U < D$ to exclude the increasing case. Note that the efficacy curve decreases monotonically when $d^U = 1$. In the increasing-plateau case, we set $d^{IP} \notin \{1, D\}$ to differentiate this relationship from the constant and increasing cases. Both $d^U$ and $d^{IP}$ were determined based only on the efficacy probabilities. Example toxicity and efficiency probability curves for all dose-response relationships are given in Figure~\ref{fig_effcurve}.

Four dose levels were considered for all simulation studies, i.e., $D = 4$. The maximum number of cohorts was set to $C_M = 10$ with a cohort size of $n_c = 3$, yielding a maximum total sample size of $N = 30$. The maximum acceptable toxicity probability was set to $p_T = 0.35$ and the minimum acceptable efficacy probability was set to $q_E = 0.25$. Common design parameters were consistently applied across all designs to enable fair comparisons. Specifically, for the BOIN-based designs, BOIN12, BOIN-ET, UBI, and STEIN, it was assumed that $\lambda_e = 0.276$ and $\lambda_d = 0.419$, given $p_T = 0.35$. For BOIN12 and uTPI, $u_2 = 40$, and $u_3 = 60$. Each trial, regardless of the design, continued until it reached the maximum total sample size or was terminated early due to the elimination of all dose levels during the trial.

We adopted the values provided in the original papers for design parameters specific to each design, such as those in each utility function. However, because our $p_T$ and $q_E$ differed from those values used in BOIN-ET, TEPI-2, and PRINTE, some of the design parameters were adjusted as follows: (1) we set $\{p_1^*, p_2^*, q_1^*, q_2^*\} = \{0.2, 0.35, 0.25, 0.6\}$ for TEPI-2 and PRINTE, which are design parameters in the truncated functions $f_1(p)$ and $f_2(q)$ for OBD selection as presented in Sections~\ref{method_TEPI-2} and \ref{method_PRINTE}; (2) we modified TEPI-2's decision table to fit our simulation studies as presented in Table S1 because the length of the toxicity and efficacy subintervals in the decision table depends on $p_T$ and $q_E$; (3) $\phi_p$ was adjusted from 0.3 to 0.35 for BOIN-ET, which yields new optimal values of $(\lambda_1, \lambda_2, \eta_1) = (0.17, 0.44, 0.48)$ through a grid search. Note that ``Table S$i$'' denotes the $i$-th table in the supplementary document.

\vspace{-0.1in}
\subsection{Generating Random Toxicity and Efficacy Probabilities}\label{prob_gen}

% Our simulation studies utilized a random generation approach to derive toxicity and efficacy probabilities. 
A total of 80,000 independent replications were carried out for each dose-response relationship, where 40,000 replications included the true OBD, while the remaining 40,000 did not. As outlined in Section~\ref{obd}, the OBD exists only if a dose $d$ meets the criteria of $p_d \leq 0.35$ and $q_d \geq 0.25$. To ensure that we generated realistic toxicity and efficacy probabilities in our simulation, we assumed that $p_d \leq p_{max} = 0.7$ and $q_d \leq q_{max} = 0.9$ for all dose levels. In every replication where the true OBD existed, we generated the toxicity and efficacy probabilities as follows:  % {\color{red} To prevent excessively similar toxicity and efficacy probability values between two dose levels, we imposed a minimum difference value of 0.05, i.e., $|p_{d+1} - p_{d}| > 0.05$ and $|q_{d+1} - q_{d}| > 0.05$ for each $d < D$. }

\vspace{-0.15in}

\begin{enumerate}
    \item We first randomly selected one dose level, $d^*$, as the OBD (among those dose levels that qualify to be the OBD) with equal probability, and then generated $p_{d^*} \sim \text{Unif}(0, p_T)$, and $q_{d^*} \sim \text{Unif}(q_E, q_{max})$.
    \vspace{-0.15in}
    \item Let $d^H$ be the dose level with the highest efficacy probability, where $q_{d^*} \leq q_{d^H}$. If more than one dose level has the highest efficacy probability, such as in the constant scenario, $d^H$ is the lowest one among these dose levels. Given $d^*$, we selected $d^H$ through a random process with equal probability among those qualified dose levels. If $d^H = d^*$, we set $p_{d^H} = p_{d^*}$ and $q_{d^H} = q_{d^*}$; if $d^H \neq d^*$, we generated $q_{d^H}\sim \text{Unif}(q_{d^*}, q_{max})$. 
    \vspace{-0.15in}
    \item After $d^*$ and $d^H$ were selected, we generated the remaining probabilities of toxicity and efficacy by following the procedures outlined in Table~\ref{tab:random}.  
    \vspace{-0.15in}
\end{enumerate}

Note that $d^*$ and $d^H$ can be different dose levels. For instance, in the increasing scenario of Figure~\ref{fig_effcurve}, $d^* = 3$ and $d^H = 4$. However, $d^*$ cannot exceed $d^H$, because if $d^* > d^H$, we would have $p_{d^*} > p_{d^H}$ under the assumption of monotonically increasing toxicity probabilities. This contradicts the definition of $d^*$. Besides, not all dose levels qualify as $d^*$ or $d^H$. For example, only dose level 1 can be $d^*$ and $d^H$ in the constant scenario. Table S3 presents all the eligible combinations of $(d^*, d^H)$ for each dose-response relationship. Additionally, Figure S1 shows example toxicity and efficacy probability curves for each combination. We also presented the algorithm for generating random probabilities of toxicity and efficacy in the replications where no OBD exists in Table S4.

% By definition, $q_{d^*}$ should not exceed $q_{d^H}$, which would imply that $d^H$ is more optimal than $d^*$ and which would contradict the definition of the OBD. If $d^* \neq d^H$, we randomly generated $q_{d^H} \sim \text{Unif}(q_{d^*}, q_{max})$.

For a fair comparison, it is crucial that the suggested OBD determined based on the given toxicity and efficacy probabilities is consistent across all designs. Therefore, in every replication, the process of generating these probabilities was repeated until they satisfied this requirement. Table S5 provides a breakdown of the frequency with which each dose level was chosen as the true OBD under each dose-response relationship. The table demonstrates that each dose level that qualifies to be the true OBD has an equal probability of being selected as the true OBD.

\vspace{-0.2in}
\subsection{Results}\label{sec_simuresult}

We evaluated seven dose finding designs using seven distinct performance metrics: 
\vspace{-0.1in}
\begin{itemize}
    \item[ - ] $p_{OBD}$: the proportion of trials that successfully select the OBD given the OBD exists; 
    \vspace{-0.1in}
    \item[ - ] $n_{OBD}$: the average number of patients assigned to the OBD given the OBD exists;
    \vspace{-0.1in}
    \item[ - ] $n_{over}$: the average number of patients assigned to dose levels with toxicity probabilities greater than $p_T + 0.1$ given the OBD exists. 
    \vspace{-0.1in}
    \item[ - ] $p_{poor}$: the proportion of trials that assign less than $20\%$ of patients to the OBD given the OBD exists;
    \vspace{-0.1in}
    \item[ - ] $p_{no}$: the proportion of trials that successfully do not select any dose level when the OBD does not exist; 
    \vspace{-0.1in}
    \item[ - ] $n_{no, over}$: the average number of patients assigned to dose levels with toxicity probabilities greater than $p_T + 0.1$ when the OBD does not exist; 
    \vspace{-0.1in}
    \item[ - ] $n_{no}$: the average number of treated patients when the OBD does not exist.
\end{itemize}
\vspace{-0.1in}
The first four metrics are applicable when the OBD exists, while the last three are used when the OBD does not exist. For each metric under each dose-response relationship, the value corresponding to the best-performing design is highlighted in bold in Table~\ref{simuresult}. The results indicate that no single design consistently outperforms the others across all metrics and scenarios.

When the OBD exists, in scenario I (Increasing), STEIN exhibits the highest value of $p_{OBD}$, whereas uTPI has the second highest $p_{OBD}$. In scenario C (Constant), UBI leads with the highest probability of accurately identifying the OBD, with PRINTE closely following UBI. When the dose-response relationship follows a unimodal (U) or increasing-plateau (IP) pattern, STEIN has the highest $p_{OBD}$. Overall, STEIN has the best performance in terms of $p_{OBD}$. We compared the selected OBD from each design to the true OBD, averaging the results across all replications for each scenario. When the selected OBD deviates from the true OBD, Table S6 shows that, on average, the percentage of cases where the selected OBD was lower than the true OBD is around 37\% for BOIN-ET, 53\% for BOIN12, 60\% for both uTPI and STEIN, and is as high as 80\% for UBI, TEPI-2, and PRINTE. This result indicates that UBI, TEPI-2, and PRINTE are more conservative compared to the other designs, and tend to select lower dose levels as the OBD. 

Regarding the metrics, $\{n_{OBD}, n_{over}, p_{poor}\}$, we found that: (1) PRINTE has the highest $n_{OBD}$ in scenarios I, U, and, IP, while uTPI has the highest $n_{OBD}$ in scenario C. (2) STEIN has the lowest $n_{over}$ in scenarios C, U, and IP, while uTPI has the lowest $n_{over}$ in scenario I. (3) In terms of $p_{poor}$, BOIN-ET performs the best in scenario I, while STEIN is the best design in scenarios C and IP, and BOIN12 is the best design in scenario U. (4) In terms of these metrics, TEPI-2 and UBI underperform compared to the other designs. For example, in scenario C, while UBI and TEPI-2 have high $p_{OBD}$ values, they also have notably low $n_{OBD}$ values and high $p_{poor}$ values. This is because both TEPI and UBI escalate to the next dose level if the current dose has a low observed toxicity probability and a high observed efficacy probability, as demonstrated in Table~\ref{TEPI-2_dec} and Table 3 of Li et al.\cite{li2020tepi}. 

When the true OBD does not exist, PRINTE consistently exhibits the highest accuracy in not choosing any dose level as the OBD, represented by $p_{no}$. TEPI-2 and UBI have smaller $p_{no}$ values than PRINTE, but they perform better than the other four designs. While BOIN-ET and STEIN have high $p_{OBD}$ values, they have much lower $p_{no}$ values than the other remaining designs. For $n_{no,over}$, STEIN is the best-performing design across all scenarios, followed by TEPI-2 and UBI. TEPI-2 and UBI have a smaller $n_{no}$ compared to other designs across all scenarios.

\vspace{-0.2in}
\subsection{Sensitivity Analysis}
To evaluate the robustness of our simulation findings, we performed five independent sensitivity analyses by changing the following: 
\vspace{-0.1in}
\begin{enumerate}
    \item [(SA1)] the number of dose levels to $D = 5$; 
    \vspace{-0.1in}
    \item [(SA2)] the maximum number of cohorts to $C_M = 15$;
    \vspace{-0.1in}
    \item [(SA3)] the correlation between the toxicity and efficacy outcomes to $\rho_1 = 0.2$;
    \vspace{-0.1in}
    \item [(SA4)] the correlation between the toxicity and efficacy outcomes at $\rho_2 = 0.4$; 
    \vspace{-0.1in}
    \item [(SA5)] the initial dose level at the beginning of the trial to dose level 2.
\end{enumerate}
\vspace{-0.1in}
In summary, the results of the first four sensitivity analyses, which used the first dose level as the initial dose, confirm the reliability and robustness of our findings in Section~\ref{sec_simuresult}. However, in SA5 which sets a different initial dose level of dose level 2, all designs had a decrease in $p_{OBD}$ and an increase in $p_{poor}$, which is particularly noticeable in the case of BOIN-ET. Detailed results of these analyses are provided in Section 7 of the supplementary document.

\vspace{-0.2in}
\section{Case Study}\label{case_sec}

Another simulation study was conducted using the CAR T-cell therapy phase I clinical trial example as illustrated in Li et al.\cite{li2020tepi} and Raje et al.\cite{raje2019anti}. The patients with relapsed or refractory multiple myeloma (RRMM) were administered a single infusion at four distinct doses during the dose escalation phase: $50\times 10^6$, $150 \times 10^6 $, $450\times 10^6 $, and $800 \times 10^6 $ CAR-positive (CAR+) T cells. The four doses were defined as dose level 1 through dose level 4. In the dose escalation phase, 21 patients were assigned using the 3+3 design, with the number of patients at each dose level $n_1 = 3$, $n_2 = 6$, $n_3 = 9$, and $n_4 = 3$. Based on the observed DLTs at each dose level, the calculated DLT rates, $\{\hat{p}_d\}_{d=1}^4$, were 0\%, 17\%, 33\%, and 67\%. The efficacy outcome was assessed using the International Myeloma Working Group (IMWG) uniform response criteria for multiple myeloma. The responders included complete response, very good partial response, and partial response. The reported response rates, $\{\hat{q}_d\}_{d=1}^4$, were 33\%, 75\%, 95\%, and 100\%\cite{li2020tepi}. The $150 \times 10^6$ and $450\times 10^6$ CAR+ T cells were selected for the dose expansion phase. We assumed $p_T = 0.35$ and $q_E = 0.25$. Due to the high values of $\hat{p}_4$ and $\hat{q}_4$, we set both $p_{max}$ and $q_{max}$ to be 1. The values of the other design parameters were the same as those specified in Section~\ref{sec:simu_settings}. 

A random simulation was conducted by generating toxicity and efficacy probabilities based on their observed rates. It was assumed that all the toxicity and efficacy probabilities have the same uninformative prior $\text{Beta}(1,1)$. The posterior distributions are given by $\text{Beta}(1 + n_d \hat{p}_d, 1 + n_d(1-\hat{p}_d))$ for toxicity and $\text{Beta}(1 + n_d \hat{q}_d, 1 + n_d(1-\hat{q}_d))$ for efficacy. We assumed that the true toxicity rates follow a monotonically increasing curve, while the true efficacy rates exhibit either an increasing (I) or an increasing-plateau (IP) pattern, with $q_{3} = q_{4}$. We assumed that the OBD exists, given the toxicity and efficacy data of dose levels 2 and 3. Based on the observed toxicity and efficacy rates, dose level 3 is the suggested OBD for all designs. However, dose level 2 had a much lower observed toxicity probability than dose level 3. It had a high observed efficacy probability, resulting in a high observed utility score and a high observed utility function value. Due to the small sample size, dose level 2 can potentially be the true OBD, as the actual toxicity and efficacy probabilities are unknown. Therefore, the true OBD was assumed to be either dose level 2 or 3. For each combination of the dose-response relationship (I or IP) and the dose level as the true OBD (dose levels 2 or 3), we conducted a simulation study with 10,000 independent replications for each of the four combinations. The approach for generating random toxicity and efficacy probabilities in this case study is detailed in Section 6 of the supplementary document. 

Table~\ref{caseselect} presents the percentage of times that each dose level is identified as the OBD in scenarios I and IP. When dose level 2 is the true OBD, all designs show a higher selection rate for dose level 2 as the OBD, compared to the selection rate of dose level 3 when it is the true OBD. When dose level 2 is the true OBD, STEIN has the highest values of $p_{OBD}$ in both scenarios, followed by BOIN-ET. However, when the true OBD is dose level 3, BOIN-ET and BOIN12 identify the OBD more accurately than the other designs. In all scenarios, UBI, TEPI-2, and PRINTE select dose level 1 as the OBD around 20\% of the time, which aligns with the conservative nature of these designs as discussed in Section~\ref{simu_sec}. However, dose level 1 is evidently not an optimal choice for the OBD, given its lower efficacy rate of 33\% compared to 75\% at dose level 2. Our results regarding TEPI-2 and UBI align with those from Li et al. \cite{li2020tepi}, who conducted a fixed simulation study based on the observed toxicity and efficacy rates. Table~\ref{caseresult} shows that BOIN-ET, BOIN12, STEIN, and uTPI outperform the other three designs in terms of $p_{OBD}$. All designs exhibit similar $n_{OBD}$, with BOIN-ET having the highest $n_{OBD}$. BOIN-ET, PRINTE, STEIN, and uTPI have lower $n_{poor}$ but higher $p_{poor}$ values than BOIN-12, UBI, and TEPI-2. Overall, STEIN has the best performance among all the designs for this case study. 

\vspace{-0.2in}
\section{Practical Implementation}\label{sec_soft}
Based on our findings in Sections \ref{simu_sec} and \ref{case_sec}, no single design consistently outperforms the others across all performance metrics in every scenario. However, we identified three designs, STEIN, PRINTE, and BOIN12, that demonstrate strong performance under specific metrics across all scenarios, serving as our guidance for practical users:

\begin{itemize}
    \item[(1)] \textbf{STEIN}: It has the highest probability of identifying the true OBD and a low likelihood of poor allocation to the OBD when the true OBD exists. STEIN provides the best overdose control, regardless of the existence of the true OBD. Based on previous trials or other prior knowledge, STEIN can be used when there is at least one admissible dose level among the doses being considered. However, STEIN is not competitive when the true OBD does not exist.
    \vspace{-0.1in}
    \item[(2)] \textbf{PRINTE}: It allocates a large number of patients to the OBD and few patients to toxic doses when the true OBD exists. In addition, PRINTE can terminate a trial without selecting any dose as the OBD with a high probability, if no dose level is admissible. This minimizes the risk of mistakenly advancing an unsuitable dose to the next trial phase. However, as a trade-off, PRINTE may expose more patients to toxic doses compared to other designs when the true OBD does not exist. 
    \vspace{-0.1in}
    \item[(3)] \textbf{BOIN12}: It is widely used as an extension of BOIN, which has been acknowledged as a "fit-for-purpose" design by the FDA. The software for implementing BOIN12 is publicly available at \href{https://trialdesign.org/one-page-shell.html#BOIN12}{www.trialdesign.org}, which features an interactive and user-friendly interface. The software also offers pre-defined language for the protocol and the statistical analysis plan, specifically tailored to the BOIN12 design. BOIN12 adopts intuitive utility scores and an RDS table for OBD selection, which can be easily implemented by practical users.
\end{itemize}
\vspace{-0.1in}
% This is particularly important when dealing with a novel compound or treatment with limited prior information.

The other designs we have examined exhibit various limitations. For instance, BOIN-ET has a low probability of terminating the trial when the true OBD does not exist. Additionally, its performance can be affected by the choice of the initial dose level. TEPI-2 and UBI underperform compared to other designs across all four performance metrics, $\{p_{OBD}, n_{OBD}, n_{over}, p_{poor}\}$, when the OBD exists under scenarios I, U, and IP. uTPI applies the utility score method, similar to BOIN12, but uTPI has more design parameters and is more complicated to implement than BOIN12.

The BOIN-ET program can be accessed at \hyperlink{https://github.com/yamagubed/boinet}{https://github.com/yamagubed/boinet}.  An R package called ``boinet'', which includes BOIN-ET, is also available. The R code of uTPI can be obtained from \hyperlink{https://github.com/haoluns/uTPI}{https://github.com/haoluns/uTPI}. We have shared our R programs for UBI, TEPI-2, PRINTE, and STEIN, which do not have either public software or R packages, at \href{https://github.com/EugeneHao/phase-I-II-designs}{https://github.com/EugeneHao/phase-I-II-designs}.

\vspace{-0.2in}
\section{Discussion}\label{discussion}
\vspace{-0.1in}
In this article, we compared seven current innovative early phase dose finding designs, namely BOIN-ET, BOIN12, TEPI-2, UBI, PRINTE, STEIN, and uTPI. BOIN-ET serves as the most direct extension of the original BOIN design. Similar to BOIN-ET, STEIN divides the toxicity-efficacy square into six distinct regions for dosing decisions, but it utilizes unimodal isotonic regression models for OBD selection. BOIN12 is another BOIN-based design that applies different utility scores for each combination of toxicity and efficacy outcomes. uTPI merges the concepts of modeling dose desirability with the chessboard design method. These four designs formulate an admissible set for dosing decisions when the current dose has an acceptable observed toxicity rate and a less-than-superb observed efficacy rate. UBI, TEPI-2, and PRINTE only incorporate data from the current dose level to make dosing decisions, with the assistance of the corresponding decision tables. UBI uses a utility function that depends on a complex set of design parameters. TEPI-2 is characterized by a simple dose finding algorithm, although the procedure for specifying its decision table is subjective. PRINTE is superior in identifying situations where the OBD is absent, given its integration of an OBD double-validation criterion.

Our simulation results indicate that no single design consistently outperforms the other designs for all performance metrics across various dose-response relationships. To account for complex dose-response relationships, we examined four distinct dose-efficacy curves. When the OBD existed, STEIN was the best design in terms of accurately selecting the OBD in scenarios I, U, and IP while UBI had the best performance in scenario C. BOIN-ET, BOIN12, and STEIN had low probabilities of poor allocation to the OBD. STEIN consistently assigned fewer patients to toxic dose levels, regardless of the existence of the true OBD. Furthermore, PRINTE had the highest probability of not selecting any dose level as the OBD when the true OBD did not exist, across all scenarios. These findings are consistent with those presented by Yamaguchi et al. \cite{yamaguchi2023optimal} and Li et al. \cite{li2023comparison}. 

Although the list of designs we evaluated overlaps with that evaluated in Shi et al. \cite{shi2024comparative}, there are some key differences. Shi et al. only considered the unimodal and increasing-plateau relationships, whereas we also considered increasing and constant relationships. Additionally, they conducted simulations only under the assumption that the true OBD exists. In contrast, we conducted additional simulation studies to compare the designs where the true OBD does not exist. Furthermore, Shi et al. focused on comparing the dose exploration algorithms of the designs and applied the same utility score approach for OBD estimation across all designs. On the other hand, we used the OBD selection approach specific to each design, as outlined in the original papers. Note that only BOIN12 and uTPI directly adopt the utility score approach for both dose exploration and OBD estimation. Other designs can underperform when using this approach for OBD estimation. For example, UBI applies a utility function for both dose exploration and OBD selection. It is not consistent to change UBI's original OBD selection method to the utility score method without adjusting its dose finding algorithm. Shi et al. showed that with the utility score approach, BOIN12 excels in OBD selection accuracy and minimizing poor dose allocation, closely followed by uTPI \cite{shi2024comparative}. However, we found that STEIN had a higher probability of identifying the true OBD than BOIN12 with its original OBD estimation approach. Both Shi et al. and our findings conclude that STEIN has better overdose control compared to other designs.

%  UBI demonstrated the best performance in terms of accurately selecting the OBD, followed by PRINTE and TEPI-2.  This can be attributed to the fact that these three designs are more conservative, typically favoring lower dose levels as the OBD compared to the other four designs.

Since the true toxicity and efficacy rates are unknown in real clinical studies, we implemented an objective random simulation approach that generates values from the assumed distributions of the toxicity and efficacy probabilities, rather than subjectively choosing potentially biased values as the true fixed rates. These distributions of toxicity and efficacy probabilities take into account the inherent uncertainty in these probabilities. In our case study with the CAR-T example, we extended our random toxicity and efficacy probability generation procedure by incorporating prior information.  In the CAR-T example, we examined scenarios with either dose levels 2 or 3 as the true OBD, considering both Increasing and Increasing-Plateau dose-efficacy relationships, of which only one scenario was analyzed through the fixed simulation studies conducted by Li et al.\cite{li2020tepi}.

The early phase designs investigated in this article have several limitations. Firstly, BOIN-ET, TEPI-2, UBI, PRINTE, STEIN, and uTPI assume that the safety and efficacy outcomes are independent, while the true relationship between safety and efficacy may be complex. However, it has been demonstrated that this independence assumption has negligible efficacy loss \cite{cai2014bayesian, guo2015teams}. We analyzed this issue via sensitivity analyses, where we considered the correlation between efficacy and toxicity probabilities, and reached the same conclusion. Secondly, designs other than BOIN12 do not consider ordinal outcomes. To the best of our understanding, gBOIN-ET\cite{takeda2022gboin} and TITE-gBOIN-ET\cite{takeda2023tite} are the two model-assisted designs for handling both ordinal toxicity and efficacy outcomes. Additionally, all designs assume that both toxicity and efficacy outcomes will be observed quickly enough to inform the dosing decisions for the subsequent patient cohort. Implementing these designs may be challenging in certain immunotherapy trials, particularly those with late-onset or pending toxicity and efficacy responses or those experiencing rapid patient accrual. To address these challenges, time-to-event (TITE) designs, such as TITE-BOIN12 \cite{zhou2022tite} and TITE-BOIN-ET \cite{takeda2020tite}, have been developed. Finally, none of the seven designs incorporate pharmacokinetic (PK) or pharmacodynamic (PD) parameters, nor do they include biomarker data in their dose finding algorithms. For instance, PKBOIN-12 is an innovative model-assisted dose finding design that integrates PK, toxicity, and efficacy to optimize dose selection \cite{sun2023pkboin}. In future work, we will examine the performance of existing designs that incorporate these factors or consider developing a new design that enhances current methodologies. Clearly, the unique challenges of oncology drug development call for more integrated dose finding approaches. These should encompass innovative study designs, advanced statistical methods, and cross-functional collaborations.

\section*{Declaration of Interest Statement}

The authors declare that they have no known competing financial interests or personal relationships that can have appeared to influence the work reported in this paper.

\section*{Data Availability}
This paper does not use any real data. The simulation code will be available on request.

\bibliographystyle{plainnat}
\bibliography{cite}

\clearpage 
\begin{table}%[hptb]
  \caption{Utility score table for binary toxicity and efficacy outcomes}\label{utility2d}
\centering
\makebox[\textwidth][c]{
    \begin{tabular}{c|c c}
\toprule
 & \multicolumn{2}{c}{Efficacy}\\
\hline
 Toxicity & Yes & No \\
\hline
 No & $u_1 = 100$ & $u_2$\\
 Yes & $u_3$ & $u_4 = 0$ \\
\bottomrule
\end{tabular}
}
\end{table}

\clearpage
\begin{table}%[hptb]
  \caption{Decision Table for BOIN-ET}\label{BOIN-ET_dec}
\centering
\makebox[\textwidth][c]{
    \begin{tabular}{c|c c c c}
\toprule
& $0 \leq \hat{p}_d \leq \lambda_1$ & $\lambda_1 < \hat{p}_d < \lambda_2$ & $\lambda_2 < \hat{p}_d \leq 1$ \\
\hline
$\eta_1 < \hat{q}_d \leq 1$ & Stay & Stay & De-escalate\\
$0\leq \hat{q}_d \leq \eta_1$ & Escalate & Escalate/Stay/De-escalate & De-escalate\\
\bottomrule
\end{tabular}
}
\end{table}

\clearpage
\begin{table*}%[htbp]
\caption{An example of a TEPI-2 decision table based on $p_T = 0.4$ and $q_E = 0.2$ }\label{TEPI-2_dec}
\centering
%\begin{threeparttable}
\begin{NiceTabular}{cccccccc}[hvlines]
\Block[c]{3-3}{ } & & & \multicolumn{5}{c}{Efficacy Rate}\\
& & & Low & Moderate & High & \multicolumn{2}{c|}{Superb} \\
& & & (0, 0.2) & (0.2, 0.4) & (0.4, 0.6) & (0.6, 0.8) & (0.8, 1)\\
\Block[c]{10-1}{Toxicity Rate} & \Block[c]{2-1}{Low} & (0, 0.08) & \cellcolor{green!50}E & \cellcolor{green!50}E & \cellcolor{green!50}E & \cellcolor{green!50}E & \cellcolor{green!50}E \\
& & (0.08, 0.16) & \cellcolor{green!50}E & \cellcolor{green!50}E & \cellcolor{green!50}E & \cellcolor{green!50}E & \cellcolor{green!50}E \\
& \Block[c]{2-1}{Moderate} & (0.16, 0.24) & \cellcolor{green!50}E & \cellcolor{green!50}E & \cellcolor{green!50}E & \cellcolor{yellow!50}S & \cellcolor{yellow!50}S \\
& & (0.24, 0.32) & \cellcolor{green!50}E & \cellcolor{green!50}E & \cellcolor{green!50}E & \cellcolor{yellow!50}S & \cellcolor{yellow!50}S \\
& High & (0.32, 0.4) & \cellcolor{red!50}D & \cellcolor{yellow!50}S & \cellcolor{yellow!50}S & \cellcolor{yellow!50}S & \cellcolor{yellow!50}S \\
& \Block[c]{5-1}{Unacceptable} & (0.4, 0.48) & \cellcolor{red!50}D& \cellcolor{red!50}D& \cellcolor{red!50}D& \cellcolor{red!50}D& \cellcolor{red!50}D\\
& & (0.48, 0.56) & \cellcolor{red!50}D& \cellcolor{red!50}D& \cellcolor{red!50}D& \cellcolor{red!50}D& \cellcolor{red!50}D\\
& & $\cdots$ & \cellcolor{red!50}D& \cellcolor{red!50}D& \cellcolor{red!50}D& \cellcolor{red!50}D& \cellcolor{red!50}D\\
& & (0.88, 0.96) & \cellcolor{red!50}D& \cellcolor{red!50}D& \cellcolor{red!50}D& \cellcolor{red!50}D& \cellcolor{red!50}D\\
& & (0.96, 1) & \cellcolor{red!50}D& \cellcolor{red!50}D& \cellcolor{red!50}D& \cellcolor{red!50}D& \cellcolor{red!50}D\\
\end{NiceTabular}
\begin{tablenotes}
%\footnotesize
\item[] \mbox{E = escalation, S = stay, D = de-escalation}
\end{tablenotes}
%\end{threeparttable}
\end{table*}

\clearpage
\begin{table*}%[htbp]
\caption{An example of a PRINTE decision table based on $p_T = 0.4$, $p_E = 0.4$, and $\epsilon = 0.05$ }\label{PRINTE_dec}
\centering
%\begin{threeparttable}
\begin{NiceTabular}{ccccccc}[hvlines]
\Block[c]{2-2}{ } & & \multicolumn{5}{c}{Efficacy Rate}\\
& & (0, 0.2) & (0.2, 0.4) & (0.4, 0.6) & (0.6, 0.8) & (0.8, 1)\\
\Block[c]{9-1}{Toxicity Rate} & (0, 0.05) & \cellcolor{green!50}E & \cellcolor{green!50}E & \cellcolor{yellow!50}S & \cellcolor{yellow!50}S & \cellcolor{yellow!50}S \\
& (0.05, 0.15) & \cellcolor{green!50}E & \cellcolor{green!50}E & \cellcolor{yellow!50}S & \cellcolor{yellow!50}S & \cellcolor{yellow!50}S \\
& (0.15, 0.25) & \cellcolor{green!50}E & \cellcolor{green!50}E & \cellcolor{yellow!50}S & \cellcolor{yellow!50}S & \cellcolor{yellow!50}S \\
& (0.25, 0.35) & \cellcolor{green!50}E & \cellcolor{green!50}E & \cellcolor{yellow!50}S & \cellcolor{yellow!50}S & \cellcolor{yellow!50}S \\
& (0.35, 0.45) & \cellcolor{green!50}E & \cellcolor{green!50}E & \cellcolor{yellow!50}S & \cellcolor{yellow!50}S & \cellcolor{yellow!50}S \\
& (0.45, 0.55) & \cellcolor{red!50}D& \cellcolor{red!50}D& \cellcolor{red!50}D& \cellcolor{red!50}D& \cellcolor{red!50}D\\
& $\cdots$ & \cellcolor{red!50}D& \cellcolor{red!50}D& \cellcolor{red!50}D& \cellcolor{red!50}D& \cellcolor{red!50}D\\
& (0.85, 0.95) & \cellcolor{red!50}D& \cellcolor{red!50}D& \cellcolor{red!50}D& \cellcolor{red!50}D& \cellcolor{red!50}D\\
& (0.95, 1) & \cellcolor{red!50}D& \cellcolor{red!50}D& \cellcolor{red!50}D& \cellcolor{red!50}D& \cellcolor{red!50}D\\
\end{NiceTabular}
\begin{tablenotes}
\item[] \mbox{E = escalation, S = stay, D = de-escalation}
\end{tablenotes}
%\end{threeparttable}
\end{table*}

\newpage

\begin{table}[h]
\caption{Generating random toxicity and efficacy probabilities under each dose-response relationship in the presence of the true OBD}
\label{tab:random}
\centering
\makebox[\textwidth][c]{
\begin{threeparttable}
\begin{tabular}{|m{2cm}|m{14cm}|}
\hline
\multicolumn{1}{|c|}{\textbf{Scenario}} & \multicolumn{1}{c|}{\textbf{Generating Random Probabilities}} \\
\hline
\multirow{2}{*}{Increasing} & \vspace{0.05in} Toxicity: generate an ascending vector of toxicity probabilities from $\text{Unif}(0, p_{d^*})$ for dose levels $\{1, \ldots, d^*-1\}$ and another ascending vector of toxicity probabilities from $\text{Unif}(p_T, p_{max})$ for dose levels $\{d^*+1, \ldots, D\}$. \vspace{0.05in}\\
\cline{2-2}
& \vspace{0.05in} Efficacy: generate an ascending vector of efficacy probabilities from $\text{Unif}(0, q_{d^*})$ for dose levels $\{1, \ldots, d^*-1\}$ and another ascending vector of efficacy probabilities from $\text{Unif}(q_{d^*}, q_{max})$ for dose levels $\{d^*+1, \ldots, D\}$. \vspace{0.05in}\\
\hline
\multirow{2}{*}{Constant} & \vspace{0.05in} Toxicity: generate an ascending vector of toxicity probabilities from $\text{Unif}( p_{d^*}, p_{max})$ for dose levels $\{2, \ldots, D\}$ where $d^* = 1$. \vspace{0.05in}\\
\cline{2-2} 
& \vspace{0.05in} Efficacy: assign $q_1$ as the efficacy probabilities for dose levels $\{2, \ldots, D\}$. \vspace{0.05in} \\
\hline
\multirow{2}{*}{Unimodal} & \vspace{0.05in} Toxicity: 
\vspace{-0.1in}
\begin{itemize}
        \item[(a)] generate an ascending vector of toxicity probabilities from $\text{Unif}(0, p_{d^*})$ for dose levels $\{1, \ldots, d^*-1\}$; 
        \vspace{-0.15in}
        \item[(b1)] if $d^* = d^H$, generate another ascending vector of toxicity probabilities from $\text{Unif}(p_{d^*}, p_{max})$ for dose levels $\{d^*+1, \ldots, D\}$; 
        \vspace{-0.15in}
        \item[(b2)] if $d^* \neq d^H$, generate another ascending vector of toxicity probabilities from $\text{Unif}(p_T, p_{max})$ for dose levels $\{d^*+1, \ldots, D\}$.
        \vspace{-0.15in}
\end{itemize} \\
\cline{2-2} 
& \vspace{0.05in} Efficacy: 
\vspace{-0.1in}
\begin{itemize}
    \item[(a)] generate an ascending vector of efficacy probabilities from $\text{Unif}(0, q_{d^*})$ for dose levels $\{1, \ldots, d^*-1\}$; 
    \vspace{-0.15in}
    \item[(b1)] if $d^* = d^H$, generate a decreasing vector of efficacy probabilities from $\text{Unif}(0, q_{d^*})$ for dose levels $\{d^*+1, \ldots, D\}$; 
    \vspace{-0.15in}
    \item[(b2)] if $d^* \neq d^H$, generate an ascending vector of   efficacy probabilities from $\text{Unif}(q_{d^*}, q_{d^H})$ for dose levels $\{d^* + 1, d^H - 1\}$ and a decreasing vector of efficacy probabilities from $\text{Unif}(0, q_{d^H})$ for dose levels $\{d^H + 1, D\}$.
    \vspace{-0.15in}
\end{itemize} \\
\hline
\multirow{2}{*}{\begin{minipage}{2cm} Increasing-\\Plateau\end{minipage}} & \vspace{0.05in} Toxicity: the procedure for generating toxicity probabilities follows the same method used in the unimodal scenario. \vspace{0.05in} \\ 
\cline{2-2} 
& \vspace{0.05in} Efficacy: 
\vspace{-0.1in}
\begin{itemize}
        \item[(a)] generate an ascending vector of efficacy probabilities from $\text{Unif}(0, q_{d^*})$ for dose levels $\{1, \ldots, d^*-1\}$;
        \vspace{-0.15in}
        \item[(b1)] if $d^* = d^H$, assign $q_{d^*}$ as the efficacy probabilities for dose levels $\{d^*+1, \ldots, D\}$; 
        \vspace{-0.15in}
        \item[(b2)] if $d^* \neq d^H$, generate an ascending vector of efficacy probabilities from $\text{Unif}(q_{d^*}, q_{d^H})$ for dose levels $\{d^*+1, \ldots, d^H-1\}$ and assign $q_{d^H}$ as the efficacy probabilities for dose levels $\{d^H+1, \ldots, D\}$.
        \vspace{-0.15in}
\end{itemize}\\
\hline
\end{tabular}
\begin{tablenotes}
%\footnotesize
\item[] (1) $d^*$: true OBD; (2) $d^H$: the lowest dose level with the highest efficacy probability; 
\item[] (3) $(p_{d^*}, q_{d^*})$: toxicity and efficacy probabilities of $d^*$; 
\item[] (4) $(p_{d^H}, q_{d^H})$: toxicity and efficacy probabilities of $d^H$; 
\item[] (5) $(p_{max}, q_{max})$: upper bounds of the generated toxicity and efficacy probabilities. 
\end{tablenotes}
\end{threeparttable}
}
\end{table}

\newpage
\begin{table}%[htbp]
\caption{Comparison of seven different performance metrics across the seven early phase designs under four scenarios for the simulation study}\label{simuresult}
\centering
\makebox[\textwidth][c]{
\begin{threeparttable}
\begin{tabular}{c|c|cccc|ccc}
\toprule
& & \multicolumn{4}{c|}{OBD Exists} & \multicolumn{3}{c}{No OBD}\\
Scenario & Design & $p_{OBD} (\%) $ & $n_{OBD}$ & $n_{over}$ & $p_{poor} (\%)$ & $p_{no} (\%)$ & $n_{no, over}$ & $n_{no}$ \\
\hline
\multirow{7}{*}{Increasing (I)}
& BOIN-ET & 64.11 & 16.29 & 2.69 & \textbf{12.46} & 20.89 & 7.80 & 27.90\\ 
& BOIN12 & 63.04 & 14.55 & 3.10 & 14.53 & 37.61 & 7.03 & 25.43 \\ 
& UBI & 59.49 & 12.99 & 4.36 & 16.66 & 45.96 & 5.03 & \textbf{22.79}\\ 
& TEPI-2 & 61.68 & 13.24 & 4.23 & 14.64 & 45.89 & 5.04 & 22.80 \\ 
& PRINTE & 64.31 & \textbf{17.12} & 2.35 & 13.44 & \textbf{52.14} & 7.95 & 25.77\\ 
& STEIN & \textbf{69.72} & 15.06 & 2.39 & 14.40 & 29.46 & \textbf{4.90} & 26.18\\ 
& uTPI & 67.52 & 15.93 & \textbf{2.29} & 17.22 & 41.01 & 7.06 & 25.56\\ 
\hline
\multirow{7}{*}{Constant (C)} 
& BOIN-ET & 62.02 & 19.68 & 2.15 & 16.90 & 22.64 & 5.97 & 28.35 \\ 
& BOIN12 & 70.02 & 18.49 & 2.10 & 10.62 & 41.97 & 4.78 & 26.38 \\ 
& UBI & \textbf{78.81} & 11.64 & 3.38 & 36.41 & 48.17 & 3.83 & \textbf{24.00}\\ 
& TEPI-2 & 75.41 & 11.93 & 3.31 & 34.90 & 48.03 & 3.87 & 24.01\\ 
& PRINTE & 77.22 & 20.13 & 2.08 & 21.30 & \textbf{60.78} & 6.64 & 26.42\\ 
& STEIN & 70.25 & 20.46 & \textbf{1.29} & \textbf{10.30} & 22.08 & \textbf{3.50} & 27.77\\ 
& uTPI & 72.89 & \textbf{20.89} & 1.71 & 14.62 & 47.73 & 4.92 & 26.06\\ 
\hline 
\multirow{7}{*}{Unimodal (U)} 
& BOIN-ET & 72.04 & 18.33 & 2.32 & 9.93 & 37.24 & 6.82 & 27.38\\ 
& BOIN12 & 72.12 & 16.74 & 2.60 & \textbf{7.79} & 36.12 & 6.42 & 25.01\\ 
& UBI & 67.72 & 13.37 & 4.01 & 18.07 & 41.03 & 5.19 & \textbf{23.42}\\ 
& TEPI-2 & 70.44 & 13.57 & 3.93 & 16.69 & 40.96 & 5.17 & \textbf{23.42}\\ 
& PRINTE & 71.06 & \textbf{18.69} & 2.19 & 13.18 & \textbf{50.40} & 6.75 & 25.29\\ 
& STEIN & \textbf{73.63} & 18.11 & \textbf{1.76} & 8.27 & 30.86 & \textbf{4.47} & 25.48\\ 
& uTPI & 72.04 & 18.61 & 1.81 & 10.66 & 38.68 & 5.83 & 25.03\\ 
\hline
\multirow{7}{*}{\begin{minipage}{2cm}
Increasing-\\Plateau (IP)
\end{minipage}}
& BOIN-ET & 67.30 & 18.04 & 2.67 & 10.51 & 22.76 & 7.24 & 27.84\\ 
& BOIN12 & 67.67 & 16.12 & 3.17 & 9.14 & 38.79 & 6.58 & 25.35\\ 
& UBI & 66.78 & 13.15 & 4.49 & 18.15 & 46.57 & 4.92 & \textbf{22.85}\\ 
& TEPI-2 & 69.61 & 13.45 & 4.37 & 16.01 & 46.50 & 4.92 & \textbf{22.85}\\ 
& PRINTE & 70.25 & \textbf{18.42} & 2.49 & 13.37 & \textbf{54.54} & 7.52 & 25.70\\ 
& STEIN & \textbf{73.64} & 17.52 & \textbf{2.22} & \textbf{8.86} & 29.81 & \textbf{4.61} & 26.09\\ 
& uTPI & 70.26 & 18.09 & 2.27 & 11.90 & 42.12 & 6.48 & 25.46\\ 
\bottomrule
\end{tabular}
\begin{tablenotes}
%\footnotesize
\item[] (1) $p_{OBD}$: OBD selection rate; (2) $n_{OBD}$: average number of patients assigned to the OBD; (3) $n_{over}$: average number of overdose patients; (4) $p_{poor}$: poor allocation proportion; and two metrics when the OBD does not exist: (5) $p_{no}$: proportion of correctly not selecting any dose; (6) $n_{no,over}$: average number of overdose patients; (7) $n_{no}$: average number of treated patients.
\end{tablenotes}
\end{threeparttable}
}
\end{table}

\clearpage
\begin{table}
\caption{The selection probability of each dose level (DL) as the OBD (\%) for all the designs under two scenarios: Increasing (I) and Increasing-Plateau (IP), and two designated true OBDs, for the case study}\label{caseselect}
\makebox[\textwidth][c]{
\begin{tabular}{c|c|cccc|cccc}
\toprule
& & \multicolumn{8}{c}{Selection Probability (\%)} \\
\hline
& & \multicolumn{4}{c|}{True OBD = DL2} & \multicolumn{4}{c}{True OBD = DL3}\\
\hline
Scenario & Design & DL1 & DL2 & DL3 & DL4 & DL1 & DL2 & DL3 & DL4 \\
\hline
\multirow{7}{*}{Increasing (I)} 
& BOIN-ET & 10.58 & 77.71 & 11.25 & 0.10 & 5.54 & 24.01 & \textbf{68.36} &  1.89 \\
& BOIN12 & 12.19 & 71.75 & 14.90 & 0.27 & 12.30 & 17.69 & 66.18 &  3.12 \\
& UBI & 26.30 & 68.42 & 3.51 & 0.01 & 20.93 & 33.83 & 43.33 &  0.38\\
& TEPI-2 & 20.97 & 73.95 & 3.33 & 0.00 & 17.77 & 30.99 & 49.45 &  0.26\\
& PRINTE & 21.57 & 74.39 & 2.54 & 0.01 & 16.37 & 33.74 & 48.72 &  0.26\\
& STEIN & 12.51 & \textbf{80.34} & 6.40 & 0.02 & 9.55 & 25.14 & 63.96 &  0.97\\
& uTPI & 16.90 & 74.33 & 7.79 & 0.07 & 11.68 & 27.78 & 58.75 &  1.15\\
\hline
\multirow{7}{*}{\begin{minipage}{2cm} Increasing-\\Plateau (IP) \end{minipage}}
& BOIN-ET & 10.86 & 81.97 &  6.77 &  0.04 & 5.57 & 24.63 & 67.90 &  1.60 \\
& BOIN12 & 9.60 & 76.61 & 12.79 &  0.13 & 11.60 & 16.43 & \textbf{68.97} &  2.26 \\
& UBI &  26.06 & 70.42 &  1.85 &  0.01 & 20.96 & 32.86 & 43.92 &  0.31 \\ 
& TEPI-2 &  18.57 & 77.38 &  2.38 &  0.02 & 17.34 & 30.71 & 49.69 &  0.33 \\ 
& PRINTE &  20.96 & 76.39 &  1.32 &  0.02 &  16.67 & 33.70 & 47.98 &  0.26 \\ 
& STEIN & 12.03 & \textbf{83.51} &  3.81 &  0.01 & 9.48 & 25.82 & 63.41 &  0.84 \\
& uTPI & 16.56 & 78.03 &  4.69 &  0.05 & 11.57 & 26.88 & 59.91 &  0.96 \\
\bottomrule
\end{tabular}
}
\end{table}

\clearpage
\begin{table}
\caption{Comparison of four performance metrics across the seven early phase designs under two scenarios: Increasing (I) and Increasing-Plateau (IP), for the case study}\label{caseresult}
\centering
\makebox[\textwidth][c]{
\begin{threeparttable}
\begin{tabular}{c|c|cccc}
\toprule
Scenario & Design & $p_{OBD} (\%) $ & $n_{OBD}$ & $n_{over}$ & $p_{poor} (\%)$ \\
\hline
\multirow{7}{*}{Inceasing (I)} 
& BOIN-ET & \textbf{73.03} & \textbf{10.74} & 0.69 & 23.13 \\ 
& BOIN12 & 68.97 & 9.43 & 1.92 & 18.75 \\ 
& UBI & 55.88 & 9.19 & 2.92 & 19.14 \\ 
& TEPI-2 & 61.70 & 9.19 & 2.91 & \textbf{18.39} \\ 
& PRINTE & 61.56 & 10.29 & \textbf{0.58} & 29.05 \\ 
& STEIN & 72.15 & 10.36 & 0.67 & 21.28 \\ 
& uTPI & 66.54 & 10.16 & 0.71 & 27.70 \\ 
% & BOIN-ET & \textbf{51.46} & 8.70 & 0.87 & 36.87 & 9.98 & 3.23 \\ 
% & BOIN12 & 49.09 & 8.41 & 1.91 & 25.66 & 17.81 & 4.18 \\  
% & UBI & 38.72 & \textbf{8.88} & 2.15 & 23.95 & 20.67 & 2.73 \\ 
% & TEPI-2 & 44.72 & 8.87 & 2.14 & \textbf{23.07} & 20.67 & 2.71 \\ 
% & PRINTE & 44.74 & 8.49 & 0.77 & 40.37 & \textbf{29.93} & 3.11 \\ 
% & STEIN & 50.32 & 8.63 & \textbf{0.74} & 32.41 & 12.59 & \textbf{1.94} \\
% & uTPI & 49.98 & 8.37 & 0.82 & 40.88 & 18.29 & 3.28\\
\hline
\multirow{7}{*}{\begin{minipage}{2cm}
Increasing-\\Plateau (IP)
\end{minipage}}
& BOIN-ET & \textbf{74.94} & \textbf{10.99} & 0.45 & 22.94 \\ 
& BOIN12 & 72.79 & 9.76 & 1.76 & \textbf{17.04}\\ 
& UBI & 57.17 & 9.25 & 2.85 & 19.93 \\ 
& TEPI-2 & 63.53 & 9.24 & 2.84 & 19.23 \\ 
& PRINTE & 62.19 & 10.46 & \textbf{0.39} & 28.78 \\ 
& STEIN & 73.46 & 10.68 & 0.46 & 20.96 \\ 
& uTPI & 68.97 & 10.47 & 0.50 & 27.02 \\ 
% & BOIN-ET & \textbf{51.41} & 8.72 & 0.61 & 38.96 & 8.37 & 2.75 \\ 
% & BOIN12 & 51.30 & 8.60 & 1.67 & \textbf{23.95} & 17.98 & 3.55 \\ 
% & UBI & 38.68 & \textbf{8.93} & 2.08 & 25.21 & 21.92 & 2.94 \\ 
% & TEPI-2 & 45.79 & 8.92 & 2.07 & 24.21 & 21.92 & 2.94 \\ 
% & PRINTE & 44.88 & 8.47 & 0.53 & 42.17 & \textbf{32.02} & 2.73 \\ 
% & STEIN & 50.73 & 8.70 & \textbf{0.55} & 34.58 & 13.79 & \textbf{2.14} \\
% & uTPI & 49.99 & 8.45 & 0.60 & 41.52 & 18.23 & 2.53\\
\bottomrule
\end{tabular}
\begin{tablenotes}
%\footnotesize
\item[] (1) $p_{OBD}$: OBD selection rate; (2) $n_{OBD}$: average number of patients assigned to the OBD; (3) $n_{over}$: average number of overdose patients; (4) $p_{poor}$: poor allocation proportion.
\end{tablenotes}
\end{threeparttable}
}
\end{table}

\clearpage

\end{document}